\documentclass[final,3p]{elsarticle}

\journal{arxiv.org}

\usepackage{algorithmic}             
\usepackage[lined,ruled,linesnumbered,commentsnumbered]{algorithm2e}
\DontPrintSemicolon
\usepackage{array}                   
\usepackage{amsmath,amssymb,amsfonts,amsthm}
\allowdisplaybreaks
\usepackage{dsfont,bm,bbm} 
\usepackage{relsize}
\usepackage{mathrsfs}
\usepackage{booktabs}
\usepackage{mathtools}
\usepackage{arydshln}
\usepackage[german,english]{babel}   
\usepackage{caption}                 
\usepackage[utf8]{inputenc}          
\usepackage[T1]{fontenc}
\usepackage{float}
\usepackage{etoolbox}
\usepackage{graphicx}                
\usepackage{lmodern}                 
\usepackage{microtype}    
\usepackage{lineno}
\usepackage{multirow}
\usepackage{calligra}
\usepackage{nicefrac}
\usepackage[caption=false]{subfig}   
\usepackage[usenames,dvipsnames]{xcolor}
\usepackage[colorinlistoftodos,prependcaption,textsize=small]{todonotes}
\usepackage[deletedmarkup=xout,authormarkup=none]{changes}
\usepackage{url} 
\usepackage[colorlinks]{hyperref}
\usepackage{cleveref}
\crefformat{equation}{(#2#1#3)}
\AtBeginDocument{%
  \hypersetup{
    citecolor=Black,
    linkcolor=Black,   
    urlcolor=Violet,
	pdfauthor={Uribe et al.}}}

\ifpdf
\hypersetup{
	pdftitle={Tomographic reconstruction with uncertain view angles},
	pdfauthor={Uribe et al.}
}
\fi

\makeatletter
\renewcommand{\@algocf@capt@plain}{above}
\makeatother

\makeatletter
\def\url@leostyle{%
\@ifundefined{selectfont}{\def\UrlFont{\sf}}{\def\UrlFont{\small\ttfamily}}}
\makeatother
\newcommand*\patchAmsMathEnvironmentForLineno[1]{%
  \expandafter\let\csname old#1\expandafter\endcsname\csname #1\endcsname
  \expandafter\let\csname oldend#1\expandafter\endcsname\csname end#1\endcsname
  \renewenvironment{#1}%
     {\linenomath\csname old#1\endcsname}%
     {\csname oldend#1\endcsname\endlinenomath}}%
\newcommand*\patchBothAmsMathEnvironmentsForLineno[1]{%
  \patchAmsMathEnvironmentForLineno{#1}%
  \patchAmsMathEnvironmentForLineno{#1*}}%
\AtBeginDocument{%
\patchBothAmsMathEnvironmentsForLineno{equation}%
\patchBothAmsMathEnvironmentsForLineno{align}%
\patchBothAmsMathEnvironmentsForLineno{flalign}%
\patchBothAmsMathEnvironmentsForLineno{alignat}%
\patchBothAmsMathEnvironmentsForLineno{gather}%
\patchBothAmsMathEnvironmentsForLineno{multline}%
}

\newcommand{\given}{\;\ifnum\currentgrouptype=16 \middle\fi|\;}
\newcommand{\suchthat}{\;\ifnum\currentgrouptype=16 \middle\fi|\;}
\newcommand{\norm}[1]{\left\Vert#1\right\Vert}
\newcommand{\abs}[1]{\left\vert#1\right\vert}

\newcommand{\dd}{\mathrm{d}}
\newcommand{\tran}{\mathsf{T}}

\newcommand{\ve}[1]{\bm{#1}}
\newcommand{\mat}[1]{\mathbf{#1}}

\DeclarePairedDelimiter\ceil{\lceil}{\rceil}

\DeclareMathOperator*{\argmin}{arg\,min}

\newcommand{\yd}{\textcolor{black}}

\newtheorem{theorem}{Theorem}[section]
\newtheorem{remark}[theorem]{Remark}

\graphicspath{{figures/}}

\begin{document}

\begin{frontmatter}	
\title{A hybrid Gibbs sampler for edge-preserving tomographic reconstruction with uncertain view angles}

\author[DTU]{Felipe Uribe\corref{cor1}} \ead{furca@dtu.dk}
\author[UoM]{Johnathan M. Bardsley} 
\author[DTU]{Yiqiu Dong} 
\author[DTU]{Per Christian Hansen} 
\author[DTU]{Nicolai A. B. Riis} 

\cortext[cor1]{Corresponding author.} 

\address[DTU]{Department of Applied Mathematics and Computer Science, Technical University of Denmark. Richard Petersens Plads, Building 324, 2800 Kgs.\ Lyngby, Denmark.}
\address[UoM]{Department of Mathematical Sciences, University of Montana. 32 Campus Drive, Missoula, MT 59812, U.S.}

\begin{abstract}
In computed tomography, data consist of measurements of the attenuation of X-rays passing through an object. The goal is to reconstruct the linear attenuation coefficient of the object's interior. For each position of the X-ray source, characterized by its angle with respect to a fixed coordinate system, one measures a set of data referred to as a \emph{view}. A common assumption is that these view angles are known, but in some applications they are known with imprecision. We propose a framework to solve a Bayesian inverse problem that jointly estimates the view angles and an image of the object's attenuation coefficient. We also include a few hyperparameters that characterize the likelihood and the priors. Our approach is based on a Gibbs sampler where the associated conditional densities are simulated using different sampling schemes\,---\,hence the term hybrid. In particular, the conditional distribution associated with the reconstruction is nonlinear in the image pixels, non-Gaussian and high-dimensional. We approach this distribution by constructing a Laplace approximation that represents the target conditional locally at each Gibbs iteration. This enables sampling of the attenuation coefficients in an efficient manner using iterative reconstruction algorithms. The numerical results show that our algorithm is able to jointly identify the image and the view angles, while also providing uncertainty estimates of both. We demonstrate our method with 2D X-ray computed tomography problems using fan beam configurations.
\end{abstract}

\begin{keyword}
computed tomography, Bayesian inverse problems, Gibbs sampler, Laplace approximation, stochastic Newton MCMC.
\vspace*{5pt}

\noindent \emph{AMS}: 60G60, 62F15, 65C05, 65R32, 65F22.
\end{keyword}
\end{frontmatter}

\section{Introduction}
In inverse problems, the aim is to compute from a set of observations the model parameters that generated them. Tomographic reconstruction is a type of inverse problem where, in the 2D case, the objective is to estimate an image of the cross-section of an object using data from a finite number of projections. 

Here, we consider X-ray \emph{computed tomography} (CT), where a projection represents the intensity loss or attenuation of the beam of X-rays as it penetrates the object. The scanning process collects many of such projections at different orientations or \emph{view angles}, and the resulting data set is referred to as sinogram. The attenuation of an X-ray is mathematically described by a line integral over a spatially varying attenuation coefficient function, representing the image of the object. The action of mapping a function to the set of its line integrals is equivalent to computing its Radon transform \cite{herman_2010}, which represents the forward CT model. The inverse problem in CT aims at finding/reconstructing the image of the object, represented in terms of the attenuation coefficient function, by combining the forward model with the sinogram data which, in practice, is always noisy.

In classical CT reconstruction, it is assumed that the geometry of the experimental set-up is perfectly known. Thus, the angular relationships between the projections are fixed \emph{a priori}. In some applications, however, the view angles might not be known with great accuracy. The uncertainty arises from unreliable angle estimates obtained with a calibration procedure or simply from the experiment configuration, e.g., due to involuntary motion of the scanned object, vibrations affecting the X-ray sources and detectors. Since the measured data introduces undesirable artifacts in the reconstructed image when wrong view angles are used, estimating the correct view angles becomes essential to obtain satisfactory reconstructions. This task is usually referred to as the angle recovery or alignment reconstitution problem \cite{basu_and_bresler_2000b}; see \cite{kolehmainen_et_al_2003, niebler_et_al_2019} for an example in dental CT. In this case, the inverse problem will not only require the estimation of the attenuation coefficient function, but also the view angles determining the geometry of the forward CT model.

The angle recovery problem is mostly solved with deterministic methods that compute the view angles directly from projection data, and then use the resulting angle orientations to reconstruct the object \cite{radermacher_1994}. Oftentimes, the geometric moments of the image are also combined with the data to guide the solution \cite{basu_and_bresler_2000b, lamberg_and_ylinen_2007}. Other deterministic methods aim at minimizing the discrepancy between the forward CT model and the data \cite{vanleeuwen_et_al_2018, yang_et_al_2005}. The angle recovery problem can also be approached using the Bayesian probabilistic framework, which offers a way to model the potential uncertainty in the reconstructed image (see, e.g., \cite{bardsley_2019, kaipio_and_somersalo_2005}). The idea is to update the \emph{prior} probability distribution of the attenuation coefficient function by including information about the forward CT model and projection data. Solving the Bayesian inverse problem consists of characterizing the updated prior information, or \emph{posterior} distribution.

Modeling the uncertainty and inferring both the attenuation coefficient and the view angles has been limited in practice, due to the large-scale nature of CT reconstruction problems. In this case, the attenuation coefficient becomes a random field, which is typically discretized pointwise on a fine rectangular grid. This complicates the use of Bayesian inference methods since one requires the exploration of a high-dimensional parameter space. Some attempts in this direction include the approaches in \cite{mallick_et_al_2006, panaretos_2009, riis_et_al_2020b}. Particularly, the CTVAE (Computed Tomography with View Angle Estimation) algorithm \cite{riis_et_al_2020b}, uses a total variation prior on the attenuation coefficients and a Gaussian prior on the view angles. However, these works characterize the Bayesian inverse problem in terms of the maximum \emph{a posteriori} probability estimator of the attenuation coefficients and there is no estimation of the posterior uncertainty.

In this paper, we formulate a Bayesian inference approach to solve the joint CT inverse problem, while also providing uncertainty estimates. Hyperparameters associated with the prior and likelihood densities are also part of the inference process, thereby defining a hierarchical Bayesian inverse problem. For the attenuation coefficients, we impose a Laplace difference prior enabling the representation of sharp edges in the image, which is often an important aspect in CT reconstruction. This prior has connections to total variation regularization, which is the standard technique in classical inverse problems for edge-preserving reconstruction (see, e.g., \cite{vogel_and_oman_1996, bardsley_2012}). For the view angles, we utilize a von Mises prior which is a $2\pi$-periodic continuous probability distribution. The resulting hierarchical posterior is simulated via the Gibbs sampler, where conditional densities are derived for each of the involved uncertain parameters. Most of these conditional distributions cannot be sampled directly, and different sampling schemes relying on Markov chain Monte Carlo (MCMC) methods are adapted for each particular Gibbs component. Hence, we consider the complete simulation approach as a hybrid Gibbs sampler \cite{roberts_and_rosenthal_1998, robert_and_casella_2004}.

Within the hybrid Gibbs sampler, the conditional distribution of the discretized attenuation coefficient is the most challenging. It defines a linear Bayesian inverse problem with non-Gaussian prior in high dimensions. For an efficient exploration of this conditional density, we define a local Laplace approximation that is utilized as a proposal distribution of a MCMC sampler. The resulting method resembles the stochastic Newton MCMC algorithm \cite{martin_et_al_2012}\,---\,the main differences being that our approach employs a quasi-Newton scheme leading to a lagged diffusivity fixed point iteration algorithm \cite{vogel_and_oman_1996}, and samples of the high-dimensional Gaussian proposal are computed efficiently via the conjugate gradient least-squares algorithm \cite{bjorck_1996}. In general, the resulting MCMC strategy is unpractical in imaging applications and we show numerically that samples from the local Laplace approximation are sufficient to characterize the attenuation coefficients and obtain satisfactory reconstructions. We test the proposed computational framework on 2D X-ray CT problems with a fan beam geometry configuration. To efficiently perform forward and back projections given a realization of the view angles, we use the matrix-free implementations in the ASTRA toolbox \cite{vanaarle_et_al_2016}. Point estimates of the posterior parameters obtained with our hybrid Gibbs sampler are compared with the solution of the CTVAE algorithm.

The organization of the paper is as follows: in \cref{sec:CT} we introduce the mathematical and computational framework, together with the assumptions and probabilistic models defining the joint CT inverse problem. The major contribution of this work is presented in \cref{sec:method} where we develop a hybrid Gibbs sampling approach to solve the resulting hierarchical Bayesian inverse problem. To estimate the high-dimensional attenuation coefficient, a sampling algorithm based on local Laplace approximations is discussed. \Cref{sec:numexp} presents and analyzes the numerical examples. The paper ends with a summary of the work and main results in \cref{sec:conclusions}.

\section{Computed tomography with uncertain view angles}\label{sec:CT}
The goal in X-ray CT reconstruction is to infer the spatially varying attenuation coefficient representing the image of an object, and we focus on \yd{the} case where the view angles from the data acquisition process are uncertain. The resulting inverse problem consists of inferring both the attenuation coefficient function and view angles. In this section, we establish a Bayesian framework to address this type of reconstruction tasks.

\subsection{Computational model}
We represent the attenuation of an X-ray (assumed to be a straight line in a two-dimensional space) using Lambert--Beer's law \cite{buzug_2008}:
\begin{equation}\label{eq:BL_law}
I_d(\theta,\tau) = I_s\exp\left(-\int_{\ell_{\theta,\tau}} x(s)\,\dd s\right) ,
\end{equation}
where $I_d(\theta,\tau)$ and $I_s$ denote the intensities at the detector and at the source, \yd{respectively,} $\ell_{\theta,\tau}$ is a line associated with angle $\theta\in [0,2\pi)$ and detector location $\tau\in \mathbbm{R}$, and $x(s)$ is the non-negative attenuation coefficient of the object at position $s\in \ell_{\theta,\tau}\subseteq\mathbbm{R}^2$. Taking logarithms in \cref{eq:BL_law} we obtain
\begin{equation}\label{eq:Radon}
b(\theta,\tau) = -\ln\left(\frac{I_d(\theta,\tau)}{I_s}\right) = \int_{\ell_{\theta,\tau}} x(s)\,\dd s  =: \mathcal{R}[x],
\end{equation}
in which $b(\theta,\tau)=-\ln({I_d(\theta,\tau)}/{I_s})$ is called the absorption (or projection). The expression \cref{eq:Radon} shows that the absorptions are given by the Radon transform $\mathcal{R}[x]$, mapping the attenuation coefficient function into the set of its line integrals.

In practice, a scanned two-dimensional object is discretized into a $N\times N$ grid, and the attenuation coefficient $x(s)$ is assumed to be constant within each grid cell (pixel). As a result, \yd{the line integrals in \cref{eq:Radon}} can be approximated as
\begin{equation}\label{eq:BL_law_discr}
\int_{\ell_{\theta,\tau}} x(s)\dd s \approx \sum_{j=1}^{d} a^{(j)}(\theta,\tau)x_j,
\end{equation}
where $a^{(j)}(\theta,\tau)$ is the length of the intersection between line $\ell_{\theta,\tau}$ and cell $j$, $x_j$ is the unknown attenuation coefficient at cell $j$, and $d=N^2$ is the total number of discretization pixels. Therefore, the discretized version of the Radon transform takes the form $\mat{A}(\ve{\theta}) \ve{x}$, where $\ve{x}\in \mathbbm{R}^d$ is the vector of attenuation coefficients, and $\mat{A}(\ve{\theta})\in\mathbbm{R}^{m\times d}$ is the so-called system matrix of intersection lengths for each of the $m$ lines and given a set of view angles $\ve{\theta}\in\mathbbm{R}^q$. The number of lines is $m= pq$, where $p$ denotes the number of detector elements and $q$ corresponds to the number of view angles.
\begin{figure}[!ht]
\centering
\includegraphics[width=0.75\textwidth]{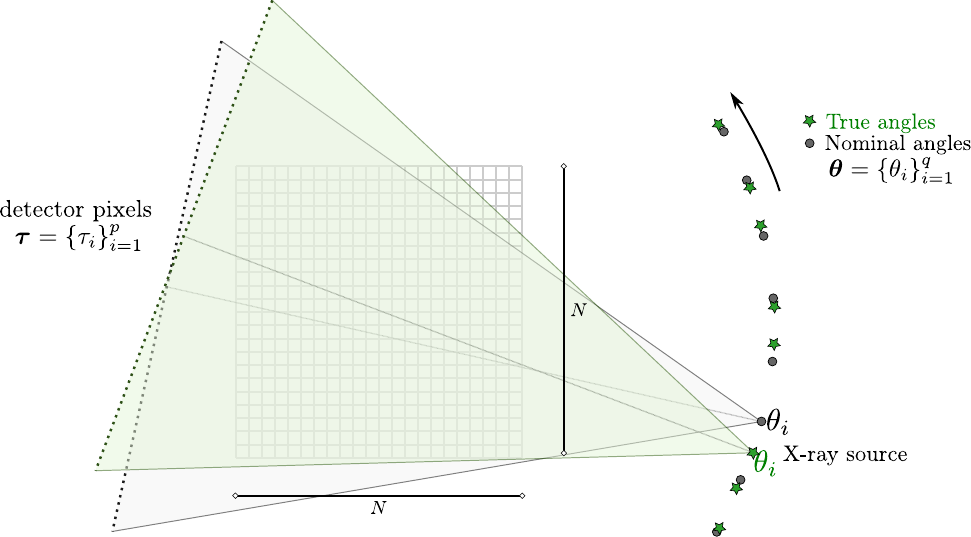}
\caption{Schematic representation of the angle recovery problem in X-ray CT: The image is computed based on a set of projections taken at assumed nominal angles (shaded gray projection). Uncertainty in these view angles lead to artifacts in the identified image. Estimation of the unknown `true' angles (shaded green projection) is a crucial step to obtain an satisfactory reconstruction.}
\label{fig:illustration}
\end{figure}

\Cref{fig:illustration} illustrates the X-ray scanning process using a fan beam configuration. At the focal point lies the X-ray source which moves on a concentric circle around the image reconstruction region. The locations of the focal point at each projection are determined by a set of view angles. The \emph{nominal angles} are predefined in the experimental set-up depending on the object geometry and radiation dosage restrictions. However, in some CT scanning settings the unknown \emph{true angles} differ from the nominal ones due to potential equipment imprecision or object/sensor movements. As a result, the computational reconstruction based on the nominal angles yields several artifacts, such as shifting, rotation or blurring of the target image object. This pathology requires the definition of a more general CT reconstruction problem in which the true and unknown view angles \yd{must be} identified to obtain a satisfactory image reconstruction.

\subsection{Inverse problem}
In X-ray tomography, the source intensity $I_s$ is typically estimated from a calibration procedure, while each intensity $I_d(\theta,\tau)$ corresponds to the line $\ell_{\theta,\tau}$. The data is then specified as the collection of projections $b=-\ln({I_d}/{I_s})$ taken during the scanning process. Hence, we consider a vector of projections $\ve{b}\in \mathcal{B}\subseteq\mathbbm{R}^{m}$ to be the noisy observational data, where $\mathcal{B}$ denotes the data space. Our aim is to utilize $\ve{b}$ to jointly infer the attenuation coefficients $\ve{x}\in \mathcal{X}\subseteq\mathbbm{R}^d$ and the view angles $\ve{\theta}\in \mat{\Theta}:=[0,2\pi)^q$. 

Let $g:\mathcal{X}\times \mat{\Theta}\to\mathcal{B}$ be the forward response operator, linking model parameters $\ve{x}$ and $\ve{\theta}$ to the data $\ve{b}$. This function is defined by the forward projection $g(\ve{x},\ve{\theta})=\mat{A}(\ve{\theta}) \ve{x}$ which corresponds to a discretized Radon transform operation. Assuming an additive measurement error, we define the inverse problem as follows:
\begin{equation}\label{eq:inv_prob}
\text{find }(\ve{x},\ve{\theta})\in\mathcal{X}\times \mat{\Theta}, \text{ such that } {\ve{b}} = \mat{A}(\ve{\theta})\ve{x} + \ve{e},
\end{equation}
where $\ve{e}\in \mathcal{B}$ is the unknown measurement noise, modeled as a realization of an independent Gaussian random vector $\ve{e}\sim \mathcal{N}(\ve{0}, \mat{\Lambda}_{\mathrm{obs}} )$ with mean zero and precision matrix $\mat{\Lambda}_{\mathrm{obs}} = \lambda\mat{I}_m$; here $\lambda$ is the precision parameter and $\mat{I}_m$ denotes the identity matrix of size $m$. We note that Gaussian noise is a good approximation to the noise present in tomographic experiments \cite{gravel_et_al_2004}; it will only deviates at the tails.

\subsection{Bayesian inverse problem}
The inverse problem \Cref{eq:inv_prob} is \yd{an ill-posed inverse problem} \cite{hansen_2010} and classical reconstruction methods must therefore employ regularization \cite{yagle_2005, vanleeuwen_et_al_2018}. Alternatively, the Bayesian approach to \Cref{eq:inv_prob} uses prior information as a mechanism of regularization \cite{kaipio_and_somersalo_2005, stuart_2010}. In this case, the unknowns are represented as random variables allowing the modeling of uncertainties due to measurement and model errors.

We consider the probability space $(\mathcal{Z}, \mathscr{B}(\mathcal{Z}), \mathds{P}_\mathcal{Z})$, with $\mathscr{B}(\mathcal{Z})= \mathscr{B}(\mathcal{X})\times\mathscr{B}(\mat{\Theta})$ a Borel $\sigma$-algebra on the product space $\mathcal{Z}=\mathcal{X}\times\mat{\Theta}$, and $\mathds{P}_\mathcal{Z}= \mathds{P}_\mathcal{X}\times\mathds{P}_{\Theta}$ a product probability measure \cite{rosenthal_2006}. We represent the uncertain parameters as a random vector with independent components $\ve{Z}$, taking values on the product space $\ve{z}=(\ve{x},\ve{\theta})\in \mathcal{Z}$. The distribution of $\ve{Z}$ has a density $\pi_{\mathrm{pr}}(\ve{z}) = \pi_{\mathrm{pr}}(\ve{x})\pi_{\mathrm{pr}}(\ve{\theta})$, where $\pi_{\mathrm{pr}}(\ve{x})=\dd \mathds{P}_\mathcal{X}/ \dd\mu$ and $\pi_{\mathrm{pr}}(\ve{\theta})=\dd \mathds{P}_{\Theta}/ \dd\mu$ are densities with respect to the Lebesgue measure $\mu$ on $\mathcal{X}$ and $\mat{\Theta}$, respectively; we refer to them as the \emph{prior} probability densities.

We assume that \yd{the measurement noise} and uncertain parameters are independent. Consider a distribution on the data space $\mathcal{B}$ with well-defined probability density $\pi(\cdot\given \ve{z})$, conditioned on a realization $\ve{z}$ of the uncertain parameters. Such density follows from the statistical model of \yd{the measurement noise} assumed in \Cref{eq:inv_prob}, linking the forward operator and data. The \emph{likelihood function} is defined from this conditional density with fixed argument equal to the measurement data, i.e., $\pi_{\mathrm{like}}(\cdot\given\ve{z}) :=\pi(\ve{b}\given \ve{z})$.

The objective of the Bayesian inverse problem is to find or characterize the \emph{posterior} probability density defined through Bayes' Theorem as \cite{kaipio_and_somersalo_2005, stuart_2010}
\begin{equation}\label{eq:Bayes}
\pi_{\mathrm{pos}}\left(\ve{z}\right):=\pi\left(\ve{z}\given \ve{b}\right) = \frac{1}{Z}  \pi_{\mathrm{like}}\left(\ve{b}\given \ve{z}\right)\pi_{\mathrm{pr}}(\ve{z}), 
\end{equation}
where $Z = \int_{\mathcal{Z}} \pi_{\mathrm{like}}\left(\ve{b}\given \ve{z}\right)\pi_{\mathrm{pr}}(\ve{z}) \dd \ve{z}$ is the normalizing constant of the posterior density, called the \emph{model evidence}. We define the Bayesian inverse problem associated to \Cref{eq:Bayes} based on the following assumptions:

\begin{itemize}
\item From the additive and independent Gaussian error model in \Cref{eq:inv_prob}, the likelihood function reads
\begin{equation}\label{eq:like}
\pi_{\mathrm{like}}(\ve{b}\given\ve{z}) = \left(\frac{\lambda}{2\pi}\right)^{\nicefrac{m}{2}} \exp\left(-\frac{\lambda}{2}\norm{\mat{A}(\ve{\theta})\ve{x}-\ve{b}}_2^2\right),
\end{equation}
where $\ve{z}=(\ve{x},\ve{\theta})$ and $\lambda\in \mathbbm{R}_{>0}$ is the precision parameter. Note that the system matrix in the forward operator is stochastic since it depends on the random vector representing the uncertain view angles.

\item For the attenuation coefficient function, we impose a zero-mean Laplace Markov random field prior \cite{bardsley_2012}. After a corresponding discretization, the field becomes a random vector with probability density
\begin{equation}\label{eq:pr_x}
\pi_{\mathrm{pr}}\left(\ve{x}\right) = \left(\frac{\delta}{2}\right)^{d} \exp\left(-{\delta}\left( \norm{\mat{D}_1\ve{x}}_1 + \norm{\mat{D}_2\ve{x}}_1 \right)\right),
\end{equation}
where $\delta\in \mathbbm{R}_{>0}$ is the inverse scale parameter and $\norm{\cdot}_1$ denotes the $\ell_1$-norm. We assume, without loss of generality, Neumann boundary conditions on the image. The horizontal and vertical first-order finite difference matrices $\mat{D}_{1},\mat{D}_2\in \mathbbm{R}^{d\times d}$ are given by
  \begin{equation}
    \mat{D}_{1} = \mat{I}_N\otimes \mat{D}\quad\text{and}\quad \mat{D}_{2} = {\mat{D}} \otimes \mat{I}_N \quad\text{with}\quad \mat{D}=\begin{bmatrix}
      -1 &  1 &        &    &   \\
         & -1 &      1 &    &   \\
         &    & \ddots & \ddots &   \\
         &    &        & -1 & 1 \\
         &    &        &    & 0 \\
    \end{bmatrix}_{N\times N},
\end{equation}
and $\otimes$ denotes the Kronecker product.

\item The view angles are modeled using a circular or periodic distribution \cite{fisher_1995}. Particularly, we assume that they are independent and impose a von Mises distribution on each of them. The resulting joint probability density can be written as
\begin{equation}\label{eq:pr_t}
\pi_{\mathrm{pr}}\left(\ve{\theta}\right) = \left(\frac{1}{2\pi I_0(\kappa)}\right)^{q} \exp\left( \kappa\ve{1}^\tran \cos(\ve{\theta}-\ve{a})\right),
\end{equation}
where $I_0(\cdot)$ is the modified Bessel function of order zero, $\kappa\in \mathbbm{R}_{>0}$ is the concentration parameter, $\ve{1}$ is a $q$-dimensional vector containing ones, and $\ve{a}\in \mathbbm{R}^q$ is the vector of mean or location parameters.
\end{itemize}

We note that the Laplace distribution has relatively heavy tails, a property that increases the probability of large increment events (jumps). This behavior facilitates the characterization of sharp edges in the unknown image, which is a desirable attribute for CT reconstruction. Particularly, it has been shown in \cite{bardsley_2012} that the model \Cref{eq:pr_x} yields piecewise constant results comparable to total variation regularization \cite{vogel_and_oman_1996}. We remark that other edge-preserving prior models exist in the literature, including: level sets \cite{dunlop_et_al_2017}, Besov space \cite{lassas_et_al_2009}, Lévy \cite{hosseini_2017}, and Cauchy differences \cite{markkanen_et_al_2019}.

The statistical models in \Cref{eq:like}, \Cref{eq:pr_x} and \Cref{eq:pr_t} require the specification of several parameters. Particularly, we focus on the precision $\lambda$, inverse scale $\delta$ and concentration $\kappa$, since their value has a considerable impact on the CT reconstruction. As a result, we also take into account the uncertainty about these values by defining them as hyperparameters in the Bayesian inverse problem. This approach is oftentimes called \emph{hierarchical regularization} (see, e.g., \cite{calvetti_and_somersalo_2008}).

\subsection{Hierarchical Bayesian inverse problem}
The Bayesian inverse problem associated to \Cref{eq:Bayes} involves the construction of more complex probability measures since \Cref{eq:like,eq:pr_x,eq:pr_t} are now conditional on random variables. Hierarchical measures can be expressed as the composition of a Markov kernel and a probability measure on each hyperparameter. Written in terms of densities, we consider the hierarchical likelihood $\pi_{\mathrm{like}}\left(\ve{b}\given\ve{x},\ve{\theta},\lambda\right) \pi_{\mathrm{hpr}}(\lambda)$, together with the hierarchical priors $\pi_{\mathrm{pr}}\left(\ve{x}\given \delta\right) \pi_{\mathrm{hpr}}(\delta)$ and $\pi_{\mathrm{pr}}\left(\ve{\theta}\given \kappa\right)\pi_{\mathrm{hpr}}(\kappa)$, where $\pi_{\mathrm{hpr}}(\lambda), \pi_{\mathrm{hpr}}(\delta)$ and $\pi_{\mathrm{hpr}}(\kappa)$ are the so-called \emph{hyperprior} probability densities. Again, we assume that these are defined with respect to the Lebesgue measure in the corresponding parameter space. 

The Bayesian inverse problem of estimating the posterior \Cref{eq:Bayes} is then reformulated as the hierarchical Bayesian inverse problem of determining the posterior density:
\begin{subequations}\label{eq:Bayes_hrc}
\begin{align}
    \pi_{\mathrm{pos}}(\ve{x},\ve{\theta},\lambda,\delta,\kappa)
      &:=\pi\left(\ve{x},\ve{\theta},\lambda,\delta,\kappa\given \ve{b}\right) \\[1mm]
      &=  \pi_{\mathrm{like}}\left(\ve{b}\given \ve{x},\ve{\theta},\lambda\right)
       \pi_{\mathrm{pr}}(\ve{x}\given\delta) \pi_{\mathrm{pr}}(\ve{\theta}\given\kappa)
       \pi_{\mathrm{hpr}}(\lambda) \pi_{\mathrm{hpr}}(\delta) \pi_{\mathrm{hpr}}(\kappa).
\end{align}
\end{subequations}

The hyperparameters are modeled with an exponential distribution (which is a special case of the gamma distribution with shape parameter equal one) \cite{gelman_2006},
\begin{equation}
\pi_{\mathrm{hpr}}(\lambda) = \beta \, \exp(-\beta\lambda) \qquad \text{and analogously for } \pi_{\mathrm{hpr}}(\delta) \text{ and } \pi_{\mathrm{hpr}}(\kappa),
\end{equation}
where $\beta$ is the rate parameter of the distributions. Following \cite{higdon_2007}, we set $\beta=10^{-4}$, such that the hyperpriors become relatively uninformative since almost equal probabilities are assigned to all events. 

In tomographic reconstruction, a point estimate of the posterior parameters is required. The two classical choices are the \emph{posterior mean} and the \emph{maximum a posteriori probability} (MAP) estimators (see, e.g., \cite{kaipio_and_somersalo_2005}). Other relevant summary statistics computed within the Bayesian framework are used to provide a measure of the posterior parameter uncertainty, e.g., marginal posterior densities, high-order statistical moments, or credible intervals. The main advantage of the MAP estimator is that oftentimes state-of-the-art optimization techniques can be used for its computation. Conversely, the estimation of the posterior mean and summary statistics involves integration; the advantage of this choice lies on the possibility to characterize the posterior parameter uncertainty. Due to the large number of random variables arising in the discretized X-ray tomography problem, the estimation of the posterior mean is typically performed via Monte Carlo methods \cite{owen_2013}. We focus on this type of approaches and propose an algorithm to efficiently estimate the posterior mean and related statistics associated with \Cref{eq:Bayes_hrc}.

\section{Hybrid Gibbs sampler}\label{sec:method}
The aim of Markov chain Monte Carlo (MCMC) is to compute samples or realizations of a Markov chain that is stationary with respect to the posterior distribution. Classical MCMC samplers include the Gibbs and Metropolis--Hastings algorithms (see, e.g., \cite{owen_2013}). Another class of MCMC methods, so-called \emph{hybrid} \cite[p.389]{robert_and_casella_2004}, simulate the model parameters simultaneously by combining different sampling schemes.

The hierarchical Bayesian inverse problem \Cref{eq:Bayes_hrc} is more suited to a hybrid procedure than a single approach. Sampling the complete set of parameters involves the exploration of the product space $\mathcal{Z}$, and also the spaces associated with the hyperparameters. The simulation can be performed more efficiently by splitting \Cref{eq:Bayes_hrc} and applying MCMC techniques tailored to each parameter space. We start by defining the target function $\pi_{\mathrm{target}}(\ve{x},\ve{\theta},\lambda,\delta,\kappa)\propto\pi_{\mathrm{pos}}(\ve{x},\ve{\theta},\lambda,\delta,\kappa)$ by replacing the assumed probabilistic models \Cref{eq:like}, \Cref{eq:pr_x} and \Cref{eq:pr_t} into \Cref{eq:Bayes_hrc} and leaving out the constant terms,
\begin{equation}\label{eq:target}
\begin{split}
\pi_{\mathrm{target}}&\left(\ve{x},\ve{\theta},\lambda,\delta,\kappa\right) = \lambda^{m/2}\, \delta^{d}\, {I_0^{-q}(\kappa)} \\
&~~~\exp\Big(-\frac{\lambda}{2}\norm{\mat{A}(\ve{\theta})\ve{x}-\ve{b}}_2^2 -{\delta}\left( \norm{\mat{D}_1\ve{x}}_1 + \norm{\mat{D}_2\ve{x}}_1 \right) + \kappa\ve{1}^\tran \cos(\ve{\theta}-\ve{a})- \beta\lambda - \beta\delta -\beta\kappa\Big).
\end{split}
\end{equation}

Performing statistical inference of the posterior \Cref{eq:Bayes_hrc} via MCMC amounts to sample the target function \Cref{eq:target}. This is a challenging task: (i) the pixel resolution in the image is often high, increasing the dimensionality of the discretized attenuation coefficients, (ii) the computational forward model is nonlinear in the view angles, and (iii) there exist nonlinear terms in the priors distributions. Therefore, we choose to  exploit the hierarchical structure of the problem to derive conditional distributions on each parameter space, instead of sampling \Cref{eq:target} directly. The conditional densities associated with the posterior \Cref{eq:Bayes_hrc} are as follows:
\begin{subequations}\label{eq:conds}
\begin{align}
\pi_1\left(\ve{x}\given \ve{\theta},\lambda,\delta\right) &\propto \exp\left(-\frac{\lambda}{2}\norm{\mat{A}(\ve{\theta})\ve{x}-\ve{b}}_2^2-{\delta}\left( \norm{\mat{D}_1\ve{x}}_1 + \norm{\mat{D}_2\ve{x}}_1 \right)\right),\label{eq:cond5}\\
\pi_2\left(\ve{\theta}\given \ve{x},\lambda,\kappa\right) &\propto \exp\left(-\frac{\lambda}{2}\norm{\mat{A}(\ve{\theta})\ve{x}-\ve{b}}_2^2 + \kappa\ve{1}^\tran\cos(\ve{\theta}-\ve{a})\right),\label{eq:cond4}\\
\pi_3\left(\lambda\given \ve{x},\ve{\theta}\right) &\propto \lambda^{m/2}\exp\left(-\lambda\left[\frac{1}{2}\norm{\mat{A}(\ve{\theta})\ve{x}-\ve{b}}_2^2 +\beta\right]\right),\label{eq:cond1}\\
\pi_4\left(\delta\given \ve{x}\right) &\propto \delta^{d}\exp\left(-\delta\left[\left( \norm{\mat{D}_1\ve{x}}_1 + \norm{\mat{D}_2\ve{x}}_1 \right) +\beta\right]\right),\label{eq:cond2}\\
\pi_5\left(\kappa\given \ve{\theta}\right) &\propto I_0^{-q}(\kappa) \exp\left(-\kappa\left[-\ve{1}^\tran\cos(\ve{\theta}-\ve{a})+\beta\right]\right).\label{eq:cond3}
\end{align}
\end{subequations}

The formulation \Cref{eq:conds} naturally induces a Gibbs-type update \cite{geman_and_geman_1984}. In particular, we consider this setting to be a three-stage Gibbs sampler in which the attenuation coefficients, the view angles and the hyperparameters define each of the stages. The Gibbs sampler generates a Markov chain $\{\ve{x}^{(j)}, \ve{\theta}^{(j)}, \lambda^{(j)}, \delta^{(j)}, \kappa^{(j)}\}_{j=1}^{n_s}$, with $n_s$ states or samples according to \Cref{algo:tsGibbs}. A word on our notation: an indexing superscript in brackets is utilized to specifically refer to samples or realizations, while an indexing subscript is used to refer to elements of the sample vector.
\begin{algorithm}[!ht]
\DontPrintSemicolon 
Initial states $\ve{x}^{(0)}, \ve{\theta}^{(0)}, \lambda^{(0)}, \delta^{(0)}, \kappa^{(0)}$\;

\For{$j = 1,\ldots, n_s$}{
	\texttt{/* Sample attenuation coefficients */}\;

	$\ve{x}^{(j)} \sim \pi_1(\cdot\given \ve{\theta}^{(j-1)}, \lambda^{(j-1)},\delta^{(j-1)})$ \;

  	\texttt{/* Sample view angles */}\;
	
	$\ve{\theta}^{(j)} \sim \pi_2(\cdot \given  \ve{x}^{(j)},\lambda^{(j-1)}, \kappa^{(j-1)})$ \;

	\texttt{/* Sample hyperparameters */}\;

	$\lambda^{(j)} \sim \pi_3\left(\cdot \given \ve{x}^{(j)},\ve{\theta}^{(j)}\right)$,~~$\delta^{(j)} \sim \pi_4\left(\cdot \given \ve{x}^{(j)}\right)$,~~$\kappa^{(j)} \sim \pi_5\left(\cdot \given  \ve{\theta}^{(j)}\right)$\;
}
\Return{$\{\ve{x}^{(j)}, \ve{\theta}^{(j)}, \lambda^{(j)},\delta^{(j)}, \kappa^{(j)}\}_{j=1}^{n_s}$}\;
\caption{{\sc Gibbs sampler \cite[p.372]{robert_and_casella_2004}}}
\label{algo:tsGibbs}
\end{algorithm}

Since most of the conditionals \Cref{eq:conds} cannot be sampled directly, the Gibbs sampler in \Cref{algo:tsGibbs} becomes hybrid. A classical sampling scheme in this category is Metropolis-within-Gibbs \cite{muller_1991}. We keep the term \emph{hybrid} (following \cite{roberts_and_rosenthal_1998, robert_and_casella_2004}) since not only Metropolis-type updates are used to characterize the components in the Gibbs sampler. In the remainder of this section, we discuss different sampling techniques to obtain draws from each conditional distribution.

\begin{remark}\label{rem:01}
The sampling schemes inside the Gibbs structure only require the generation of one sample for every Gibbs iteration. Thus, sampling each conditional requires in principle a single simulation. This is because the Markov chains associated to the individual conditionals are not required to reach stationarity at each Gibbs iteration. The reasons for this are discussed in \cite[p.393]{robert_and_casella_2004}, and are mainly related to structure and converge properties of the Gibbs sampler  \cite{agapiou_et_al_2014, roberts_and_rosenthal_1998}. Nevertheless, some authors suggest that performing a few extra within-Gibbs iterations can be beneficial to accelerate the convergence of the Gibbs chain and to reduce correlation in the resulting posterior samples (see, e.g., \cite{gamerman_and_lopes_2006}).
\end{remark}

\subsection{Sampling the attenuation coefficients}\label{subsec:locallap}
The conditional $\pi_1$ in \Cref{eq:cond5} defines a Bayesian inverse problem with Gaussian likelihood and Laplace difference prior. In this case, the view angles are fixed and  $\mat{A}(\ve{\theta})=\mat{A}$. The task of sampling $\ve{x}$ from $\pi_1$ faces two main difficulties: (i) the $\ell_1$-norm components in the Laplace prior make the inverse problem nonlinear, and (ii) the high-dimensional parameter space arising from the discretized attenuation coefficient field makes the sampling of $\pi_1$ prohibitive. To address these issues, we develop a method exploiting the ideas of the \emph{Hessian-based} and \emph{stochastic Newton} MCMC methods proposed in \cite{qi_and_minka_2002, martin_et_al_2012}. Our approach is tailored to Bayesian inverse problems that utilize Laplace difference or total variation priors.

The central idea is to represent $\pi_1$ in terms of an asymptotic Gaussian distribution. This approach is generally known as the \emph{Laplace approximation} \cite{tierney_and_kadane_1986}, and it relies on a Taylor series expansion of the logarithm of the target density around its mode. Despite it is possible to include higher order terms in the expansion, standard Gaussian approximations retain up to the second order (i.e., up to Hessian information). The Laplace approximation is suitable as long as the target distribution is similar enough to a Gaussian. This is a reasonable approximation in our case because the forward operator is linear on $\ve{x}$ and the Laplace difference prior is directly related to total variation regularization\,---\,it has been shown in \cite{lassas_and_siltanen_2004} that this prior is asymptotically Gaussian as the resolution increases.

Hence, we seek an approximation to $\pi_1$ in terms of a Gaussian density ${\pi}_{\mathrm{L}}(\ve{x})=\mathcal{N}(\ve{x}; \ve{\mu}_{\mathrm{L}},\mat{\Lambda}_{\mathrm{L}})$, where the mean vector $\ve{\mu}_{\mathrm{L}}$ is computed as the MAP estimator of ${\pi}_1$ and the precision matrix $\mat{\Lambda}_{\mathrm{L}}$ is equal to the Hessian of negative logarithm of ${\pi}_1$ \cite{schillings_et_al_2020}. We discuss in the following a procedure to estimate these quantities.

Due to the non-differentiability of the absolute value in the $\ell_1$-norm components of $\pi_1$, a smooth representation is imposed. The $\ell_1$-norm for the increment vectors in the horizontal and vertical directions can be written as
\begin{equation}
\norm{\mat{D}_1\ve{x}}_1 = \sum_{i=1}^d\abs{[\mat{D}_1\ve{x}]_i}
    \qquad \hbox{and} \qquad
    \norm{\mat{D}_2\ve{x}}_1 = \sum_{i=1}^d\abs{[\mat{D}_2\ve{x}]_i},
\end{equation}
respectively, where $[\cdot]_i$ denotes the $i$th element of the vector. We follow \cite[p.131]{vogel_2002} to define the smooth approximation of the absolute value $\abs{t}\approx \frac{1}{2}\psi(t)$, where the function $\psi(t) = 2\sqrt{t^2+\varepsilon}$ and the constant $0<\varepsilon\ll 1$ controls the approximation. With this representation, the negative logarithm of $\pi_1$ or \emph{cost function}, becomes
\begin{subequations}
\begin{align}\label{eq:costfunc}
J(\ve{x}) &= \frac{\lambda}{2}\norm{\mat{A}\ve{x}-\ve{b}}_2^2 + \frac{\delta}{2}\left[\sum_{i=1}^{N-1}\sum_{j=1}^{N} \psi(x_{i+1,j}-x_{i,j}) + \sum_{i=1}^{N}\sum_{j=1}^{N-1} \psi(x_{i,j+1}-x_{i,j}) \right]\\
&= \frac{\lambda}{2}\norm{\mat{A}\ve{x}-\ve{b}}_2^2 + {\delta}\left[\sum_{i=1}^{d} \sqrt{([\mat{D}_1\ve{x}]_i)^2+\varepsilon} + \sum_{i=1}^{d} \sqrt{([\mat{D}_2\ve{x}]_i))^2+\varepsilon} \right].
\end{align}
\end{subequations}

In \Cref{eq:costfunc}, with a slight abuse of notation, $x_{i,j}$ denotes the $(i,j)$ element when $\ve{x}$ is reshaped into an $N\times N$ array. For the gradient vector and Hessian matrix of $J(\ve{x})$ we have, respectively,
\begin{equation}\label{eq:grad_hess}
\nabla J(\ve{x}) = \lambda\mat{A}^\tran(\mat{A}\ve{x}-\ve{b}) + \delta \mat{L}(\ve{x})\ve{x} \quad\text{and}\quad 
\nabla^2 J(\ve{x}) \approx \widetilde{\nabla}^2J(\ve{x})=\lambda\mat{A}^\tran\mat{A} + \delta \mat{L}(\ve{x});
\end{equation}
here we have neglected a term $\delta\nabla\mat{L}(\ve{x})\ve{x}$ from the Hessian to obtain $\widetilde{\nabla}^2J(\ve{x})$, and the matrix $\mat{L}(\ve{x})\in \mathbbm{R}^{d\times d}$ is given by
\begin{equation}\label{eq:L}
\mat{L}(\ve{x}) = \mat{D}_1^\tran \mat{W}_1(\ve{x})\mat{D}_1 + \mat{D}_2^\tran \mat{W}_2(\ve{x})\mat{D}_2,
\end{equation}
where the diagonal weight matrices $\mat{W}_1(\ve{x}),\mat{W}_2(\ve{x})\in \mathbbm{R}^{d\times d}$ are equal to
\begin{equation}\label{eq:weights}
\mat{W}_1(\ve{x}) =~\text{diag}(1/\sqrt{(\mat{D}_1\ve{x})^2+\varepsilon}) \qquad\text{and}\qquad \mat{W}_2(\ve{x})= \text{diag}(1/\sqrt{(\mat{D}_2\ve{x})^2+ \varepsilon}),
\end{equation}
with \yd{the division,} the squaring and the square root taken elementwise.

Since we use an approximation of the Hessian in \Cref{eq:grad_hess}, the MAP estimator of $\pi_1$ can be computed using a quasi-Newton scheme that minimizes \Cref{eq:costfunc}. The quasi-Newton step can be written as
\begin{equation}\label{eq:LDFP} 
\ve{x}^{(k+1)} = {\ve{x}^{(k)}} - \left[\widetilde{\nabla}^2J({\ve{x}^{(k)}})\right]^{-1}\nabla J({\ve{x}^{(k)}}) =  (\lambda\mat{A}^\tran\mat{A} + \delta \mat{L}({\ve{x}^{(k)}}))^{-1}\lambda\mat{A}^\tran\ve{b};
\end{equation}
iterating through \Cref{eq:LDFP} leads to a method known as the \emph{lagged diffusivity fixed point} algorithm \cite{vogel_and_oman_1996}, also known as half-quadratic regularization \cite{chan_and_mulet_1997}. This method estimates the MAP of a nonlinear inverse problem with Laplace difference prior, by iteratively computing MAP estimators of a sequence of Gaussian linear inverse problems.

We are now in a position to define the Laplace approximation. This Gaussian density has precision matrix given by the approximate Hessian $\widetilde{\nabla}^2J(\ve{x})$, and mean vector computed from the quasi-Newton step, i.e.,
\begin{subequations}
\begin{align}\label{eq:Laplaceapprox} 
 {\pi}_{\mathrm{L}}(\ve{x};\ve{x}^{(k)}) &= \frac{\det(\mat{\Lambda}_{\mathrm{L}}({\ve{x}^{(k)}}))^{\nicefrac{1}{2}}}{(2\pi)^{\nicefrac{d}{2}}}\exp\left(-\frac{1}{2}\norm{\ve{x}-\ve{\mu}_{\mathrm{L}}({\ve{x}^{(k)}})}_{\mat{\Lambda}_{\mathrm{L}}({\ve{x}^{(k)}})}^2\right)\\\label{eq:Laplaceparams} 
\mat{\Lambda}_{\mathrm{L}}(\ve{x}^{(k)}) &= \lambda\mat{A}^\tran\mat{A} + \delta \mat{L}({\ve{x}^{(k)}}), \qquad \ve{\mu}_{\mathrm{L}}(\ve{x}^{(k)}) =  \mat{\Lambda}_{\mathrm{L}}^{-1}(\ve{x}^{(k)})\lambda\mat{A}^\tran\ve{b}.
\end{align}
\end{subequations}

Note that this Laplace approximation is \emph{local}, with its parameters depending on the state $\ve{x}^{(k)}$. We point out that a learning rate can be introduced in the quasi-Newton step \Cref{eq:LDFP} (i.e., in the mean of the Laplace approximation) to account for cases where the target distribution is not well approximated by a Gaussian (see, e.g., \cite{qi_and_minka_2002}).

Recall that the objective is to draw samples from $\pi_1$. This can be performed using the Metropolis--Hastings (MH) algorithm with proposal distribution $q(\ve{x}\given\ve{x}^{(k)})$ equal to the Laplace approximation, i.e., $q(\ve{x}\given\ve{x}^{(k)})= {\pi}_{\mathrm{L}}(\ve{x};\ve{x}^{(k)})$. Since the Laplace approximation updates its mean with the quasi-Newton direction, samples are more likely to be drawn from a high probability region of $\pi_1$. This idea relates to the Hessian-based \cite{qi_and_minka_2002} and stochastic Newton \cite{martin_et_al_2012} MCMC methods where, given $\ve{x}^{(k)}$, one draws a candidate sample $\ve{x}^\star$ from $q(\cdot\given\ve{x}^{(k)})$ and then passes it through the accept-reject mechanism of MH that has acceptance probability
\begin{equation}\label{eq:MALLA}
\alpha_{\mathrm{MH}}(\ve{x}^{(k)},\ve{x}^\star) = \min\left(1,\frac{{\pi}_{1}(\ve{x}^\star)q(\ve{x}^{(k)}\given\ve{x}^\star)}{{\pi}_{1}(\ve{x}^{(k)})q(\ve{x}^\star\given\ve{x}^{(k)} )}\right) =\min\left(1,\frac{{\pi}_{1}(\ve{x}^\star){\pi}_{\mathrm{L}}(\ve{x}^{(k)};\ve{x}^\star)}{{\pi}_{1}(\ve{x}^{(k)}){\pi}_{\mathrm{L}}(\ve{x}^\star;\ve{x}^{(k)})}\right).
\end{equation}
 
Evaluating the acceptance probability \Cref{eq:MALLA} requires inverting and computing the determinant of two precision matrices $\mat{\Lambda}_{\mathrm{L}}({\ve{x}^\star})$ and $\mat{\Lambda}_{\mathrm{L}}({\ve{x}^{(k)}})$, which becomes computationally infeasible as $d$ increases. However, since the simulation of $\ve{x}$ is embedded in a Gibbs structure, the MCMC correction can be reduced to an \emph{independence sampler} (IS) \cite{owen_2013}. This is because at each Gibbs step, we are proposing from a local Laplace approximation given the state $\ve{x}^{(k)}$,  defined from the previous Gibbs iteration. Hence, we only have to consider one Hessian at the current Gibbs step, i.e., $\mat{\Lambda}_{\mathrm{L}}({\ve{x}^{(k)}})$. In this case, the acceptance probability simplifies to
\begin{equation}\label{eq:IS}
\alpha_{\mathrm{IS}}(\ve{x}^{(k)},\ve{x}^\star) = \min\left(1,\frac{{\pi}_{1}(\ve{x}^\star)q(\ve{x}^{(k)})}{{\pi}_{1}(\ve{x}^{(k)})q(\ve{x}^\star )}\right) =\min\left(1,\frac{{\pi}_{1}(\ve{x}^\star){\pi}_{\mathrm{L}}(\ve{x}^{(k)};\ve{x}^{(k)})}{{\pi}_{1}(\ve{x}^{(k)}){\pi}_{\mathrm{L}}(\ve{x}^\star;\ve{x}^{(k)})}\right).
\end{equation}

In stochastic Newton MCMC, the computations associated to the Hessian matrix are performed by building a suitable low-rank approximation. The low-dimensional structure is represented by the parameter subspace where data is more informative relative to the prior. In our case, however, finding the low-rank approximation is not straightforward since not only the Hessian is updated at every Gibbs iteration, but also the prior is heavy-tailed and non-Gaussian. 

In general, adjusting the samples from the local Laplace approximation via \Cref{eq:MALLA} or \Cref{eq:IS} makes the problem unpractical in imaging applications. Hence, a simpler algorithm is to consider an \emph{unadjusted Laplace approximation}, which assumes $\pi_1\left(\ve{x}\given \ve{\theta},\lambda,\delta\right)\approx{\pi}_{\mathrm{L}}(\ve{x};\ve{x}^{(k)})$ around a state $\ve{x}^{(k)}$, and simulates directly from \Cref{eq:Laplaceapprox} without using an accept-reject step. Despite the resulting samples produce a biased approximation to $\pi_1$, this strategy is justified not only in the discretization limit (since $\pi_1$ is asymptotically Gaussian), but also by imposing a stringent smoothing parameter $\varepsilon$ (since we retrieve $\pi_1$ as $\varepsilon\to 0$). We comment on the application of the MCMC steps \Cref{eq:MALLA} and \Cref{eq:IS} in \Cref{rem:02}. 

\yd{The approach discussed in this section is embedded in a Gibbs structure. Since the unadjusted Laplace approximation is employed, we only require a single within-Gibbs iteration. We set the current state in the local Laplace approximation as $\ve{x}^{(j)}$, which is equal to the values of the attenuation coefficients at the previous Gibbs step. Thus, the final process to our sampling scheme is to simulate the local Laplace approximation \Cref{eq:Laplaceapprox} given $\ve{x}^{(j)}$}. We use the fact that a draw from the Gaussian approximation \Cref{eq:Laplaceapprox} can be obtained as $\ve{x}^\star=\ve{\mu}_{\mathrm{L}}({\ve{x}^{(j)}}) + \mat{\Lambda}^{-1/2}_{\mathrm{L}}({\ve{x}^{(j)}})\ve{\xi}$, where $\ve{\xi}\sim\mathcal{N}(\ve{0},\mat{I}_d)$ is a standard Gaussian random vector. \yd{According to the definitions of $\mat{A}$ and $\mat{L}$, the intersection of their null spaces is trivial, therefore $\mat{\Lambda}_{\mathrm{L}}$ is positive definite. Then, for} a symmetric and positive-definite covariance matrix $\mat{\Lambda}_{\mathrm{L}}^{-1}({\ve{x}^{(j)}})$, the task of sampling a Gaussian random vector $\ve{x}^\star$ can be written in terms of the linear system of equations:
\begin{equation}\label{eq:normal_CG}
[\lambda\mat{A}^\tran\mat{A} + \delta\mat{L}(\ve{x}^{(j)})]\ve{x}^\star = \lambda\mat{A}^\tran\ve{b}  +  \mat{\Lambda}_{\mathrm{L}}^{-1/2}({\ve{x}^{(j)}})\ve{\xi},
\end{equation}
which can be solved using the conjugate gradient algorithm (see, e.g., \cite{bardsley_2019, orieux_et_al_2015}). However, a more stable and efficient solution to \Cref{eq:normal_CG} can be computed by re-writing the problem as a least-squares task. Hence, given the current state $\ve{x}^{(j)}$, we draw a sample $\ve{x}^\star$ by solving the least-squares problem:
\begin{equation}\label{eq:CGLS}
    \argmin_{\ve{x}^\star\in \mathcal{X}}
    \| \mat{M}(\ve{x}^{(j)}) \ve{x}^\star - \ve{y} \|_2
    \quad \hbox{with} \quad \mat{M}(\ve{x}^{(j)}) =
    \begin{bmatrix}
      \sqrt{\lambda}\mat{A}\\
      \sqrt{\delta} \mat{W}_1^{\nicefrac{1}{2}}(\ve{x}^{(j)})\mat{D}_{1}\\
      \sqrt{\delta} \mat{W}_2^{\nicefrac{1}{2}}(\ve{x}^{(j)})\mat{D}_{2}
    \end{bmatrix} , \quad \ve{y} =
    \begin{bmatrix}
      \sqrt{\lambda}\ve{b} \\
      \ve{0}_{d}\\
      \ve{0}_{d}
    \end{bmatrix} + \overline{\ve{\xi}} ,
\end{equation}
where $\overline{\ve{\xi}}\sim\mathcal{N}(\ve{0},\mat{I}_{m+2d})$. We use the \emph{conjugate gradient least-squares} (CGLS) algorithm \cite[Ch.~7]{bjorck_1996} to efficiently solve \Cref{eq:CGLS}, and we note that other iterative methods are also available. All these algorithms require a pair of forward and backward model computations (i.e., multiplications with $\mat{A}$ and $\mat{A}^\tran$), per iteration. When used for sampling, a maximum number of iterations $n_{\mathrm{cgls}}$ is imposed as stopping criterion. We point out that a similar approach is adopted in the cyclic optimization algorithm for MAP estimation proposed in \cite{calvetti_and_somersalo_2008}.

\begin{remark}\label{rem:02}
We tested the MCMC adjustment step to the samples proposed by the local Laplace approximation on a small scale example. When performing one within-Gibbs iteration using the MH acceptance probability in \Cref{eq:MALLA}, the samples are always rejected. While using the IS with acceptance probability \Cref{eq:IS}, the samples are always accepted. This is related to the fact that the target conditional densities are changing with every Gibbs iteration. For the MH step, the proposal distributions are `centered' on samples from two different conditional densities (one at the previous step and one at the current), this makes the proposal ratio excessively small and dominates the acceptance probability. Conversely, for the IS step, the proposals are centered at the same state, and the proposed sample is typically on a higher probability region compared to the sample in the previous Gibbs step. We noted that the acceptance probability becomes smaller, if one increases the number of within-Gibbs iterations for the IS step. Recall, however, that drawing a single sample is in principle sufficient for the Gibbs sampler. As a result, the attenuation coefficients are simulated using the unadjusted Laplace approximation, which is equivalent to an IS setting with one within-Gibbs iteration. 
\end{remark}

\subsection{Sampling the view angles}\label{sec:samp_theta}
The conditional $\pi_2$ in \Cref{eq:cond4} defines a nonlinear inverse problem with Gaussian likelihood and von Mises prior. Despite the inverse problem is still linear in $\ve{x}$, the matrix $\mat{A}(\ve{\theta})$ is now stochastic and new forward projections of $\ve{x}$ are computed for every realization of the view angles.

When many view angles are used, standard MH leads to slow convergence rates when sampling the full view angle random vector. Another difficulty is that \Cref{eq:cond4} is potentially sharp at every mode, since small changes in the view angles are expected in practice. Given that the random vector of view angles has independent components, a suitable strategy is to perform component-wise updates where one variable (or a block of variables) are sampled individually (see, e.g., \cite{johnson_et_al_2013}). Particularly, we utilize a standard \emph{single-component Metropolis algorithm} with random-walk proposals. The method is essentially a Gibbs sampler with the conditional densities defined by marginalization of certain parameters components. Since we assume independent view angles, the sub-conditionals associated with the density \Cref{eq:cond4} are simply defined by fixing angle coordinates. The approach of inferring one component of $\ve{\theta}$ at a time also benefits from the fact that the forward projections in the computational CT model can be easily split for each individual angle. 

Let us re-arrange the projection data in terms of a \emph{sinogram} consisting of the two-dimensional array $\mat{S}\in\mathbbm{R}^{p\times q}$ with $\ve{b}=\mathrm{vec}(\mat{S})$ (column stacking operation). The forward projection performed along a single angle $\mat{A}(\theta_i)\ve{x}$, where \yd{$\mat{A}(\theta_i)\in\mathbbm{R}^{p\times d}$} is a block of $\mat{A}$, corresponds to the data vector $\ve{s}_i\in\mathbbm{R}^{p}$ equal to the $i$th column of the matrix $\mat{S}$.

We write the conditional $\pi_2$ as the product of $q$ component densities:
\begin{equation}\label{eq:pi2_decomp}
 \pi_2(\ve{\theta}\given \ve{x},\lambda,\kappa) \propto  \prod_{i=1}^q \pi\big({\theta}_i\given \ve{x},\lambda,\kappa\big), \quad \pi\big({\theta}_i\given \ve{x},\lambda,\kappa\big) \propto \exp\left(-\frac{\lambda}{2}\norm{\mat{A}\big({\theta}_i\big)\ve{x}-\ve{s}_i}_2^2 +  \kappa \cos\big({\theta}_i-{a}_i\big)\right).
\end{equation}

Given a state $\ve{\theta}^{(k)}= \bigl[{\theta}_1^{(k)},{\theta}_2^{(k)},\ldots,{\theta}_q^{(k)}\bigr]^\tran$, samples from $\pi_2$ are drawn component by component, as follows
\begin{equation}\label{eq:conds_CW}
\begin{split}
{\theta}_{1}^{(k+1)} &\sim \pi_2\left({\theta}\given  \ve{x},\lambda,\kappa,  \bigl[{\theta}_{2}^{(k)},{\theta}_{3}^{(k)},\ldots,{\theta}_{q}^{(k)}\bigr]\right),\\
{\theta}_{2}^{(k+1)} &\sim \pi_2\left({\theta}\given \ve{x},\lambda,\kappa,
  \bigl[ {\theta}_{1}^{(k+1)},{\theta}_{3}^{(k)},\ldots,{\theta}_{q}^{(k)}\bigr]\right), \\
&~~~~~\vdots \\
{\theta}_{q}^{(k+1)} &\sim \pi_2\left({\theta}\given \ve{x},\lambda,\kappa,
  \bigl[ {\theta}_{1}^{(k+1)},{\theta}_{2}^{(k+1)},\ldots,{\theta}_{q-1}^{(k+1)}\bigr]\right).
\end{split}
\end{equation}

The sub-conditionals \Cref{eq:conds_CW} are simulated individually with a random walk Metropolis algorithm. Specifically, drawing a sample $\ve{\theta}^{(k+1)}$ involves cycling through all sub-conditionals \Cref{eq:conds_CW}, proposing values ${\theta}^{\star}\sim \mathcal{N}({\theta}_{i}^{(k)}, \sigma^2)$ and then accepting/rejecting each proposal according to the Metropolis acceptance probability
\begin{equation}\label{eq:alpha_theta}
\alpha\big({\theta}_i^{(k)},{\theta}^\star\big) =\min\left(1,\frac{{\pi}({\theta}^{\star}\given  \ve{x},\lambda,\kappa\big)}{{\pi}\big({\theta}_i^{(k)}\given  \ve{x},\lambda,\kappa\big)}\right); 
\end{equation}
note that the target densities \Cref{eq:conds_CW} can be simplified given the structure imposed in the left-hand side of \Cref{eq:pi2_decomp}. Hence, only evaluation of $\pi_2$ along the $i$th component is required to compute \Cref{eq:alpha_theta}, i.e., the individual component densities in the right-hand side of \Cref{eq:pi2_decomp}.
\begin{figure}[!ht]
\centering
\includegraphics[width=0.9\textwidth]{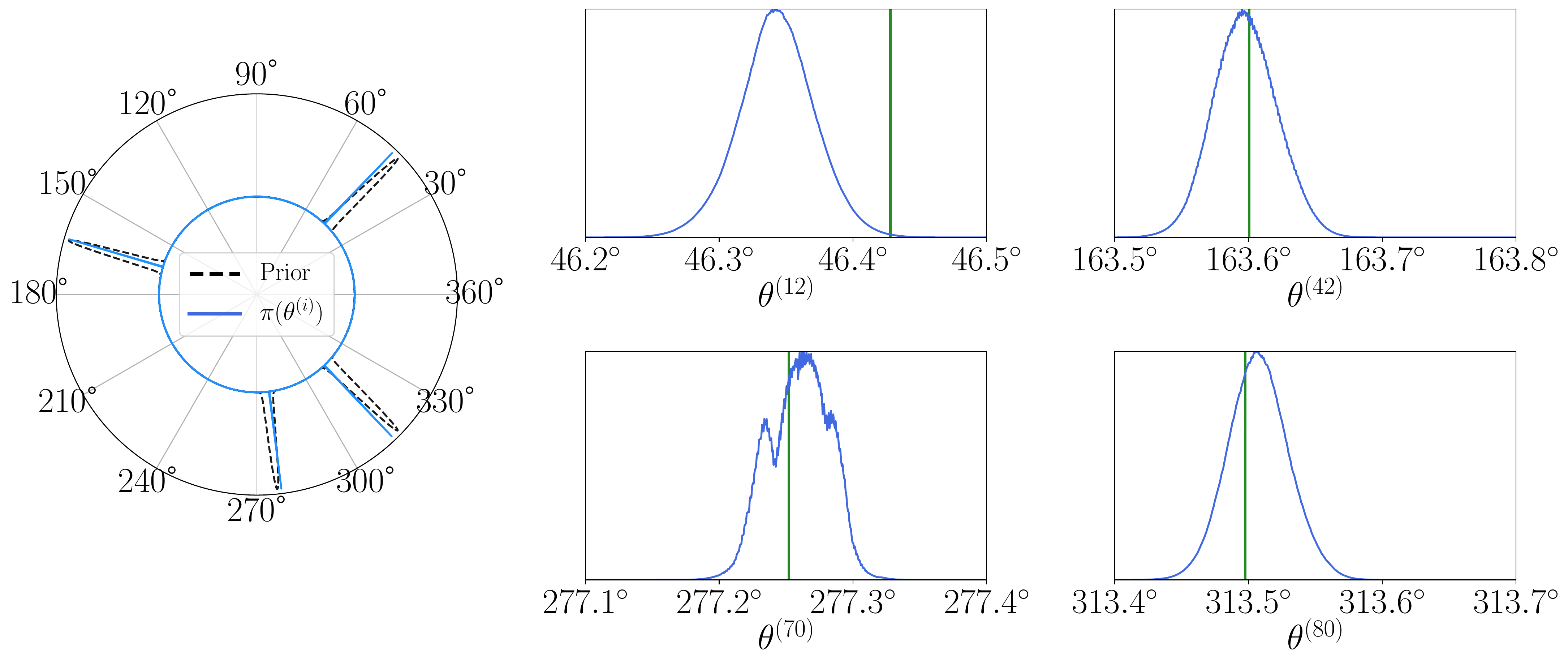}
\caption{Selected component densities \Cref{eq:pi2_decomp} from the conditional $\pi_2$. The values of $\ve{x}$, $\lambda$ and $\kappa$ are assumed known. Left: von Mises prior with the respective component densities. Right: zoom-in on component densities and true angles shown as solid green lines.}
\label{fig:component_thetas}
\end{figure}

The proposal scale $\sigma$ can be adapted for each component using the methods proposed in \cite{haario_et_al_2001, andrieu_and_thoms_2008}. However, the sub-conditionals \Cref{eq:conds_CW} are sometimes multimodal. In our experiments, we observe that this property seems to be induced by the specific implementation of the forward CT operator. A similar behavior is reported in \cite{burger_et_al_2020}, where the target functions depending on the view angles have multiple local optima. \Cref{fig:component_thetas} illustrates this phenomenon at selected single-component conditional densities \Cref{eq:pi2_decomp} (for \yd{the 50-grain phantom} in \Cref{subsec:grains}). These densities are computed for illustrative purposes assuming that the true values of the attenuation coefficients, noise precision and angle concentration parameters are known. We see that some of the densities are multimodal and relatively noisy, which complicates the application of traditional MCMC adaptation schemes. Hence, when using the component-wise Metropolis algorithm, the individual Gaussian proposal scalings are fixed. We found via parameter studies that $5\%$ of the nominal view angles spread is a good heuristic for these proposal scales.

\subsection{Sampling the hyperparameters}\label{subsec:hyper}
The conditional $\pi_3$ in \Cref{eq:cond1} can be written in closed-form due to the conjugate relation between the gamma and Gaussian distributions. The resulting density is gamma-distributed defined as \cite{bardsley_2019, higdon_2007}
\begin{equation}\label{eq:cond_hyper_3} 
\pi_3\left(\lambda\given \ve{x},\ve{\theta}\right) = \frac{\beta_{\lambda}^{\alpha_{\lambda}}}{\Gamma(\alpha_{\lambda})}\lambda^{\alpha_{\lambda}-1}\exp(-\beta_{\lambda}\lambda),\quad \text{ with } \alpha_{\lambda} = \frac{m}{2}+1 \text{ and }\beta_{\lambda}= \frac{1}{2}\norm{\mat{A}(\ve{\theta})\ve{x}-\ve{b}}_2^2 + \beta.
\end{equation}

\begin{figure}[!ht]
\centering
\includegraphics[width=0.55\textwidth]{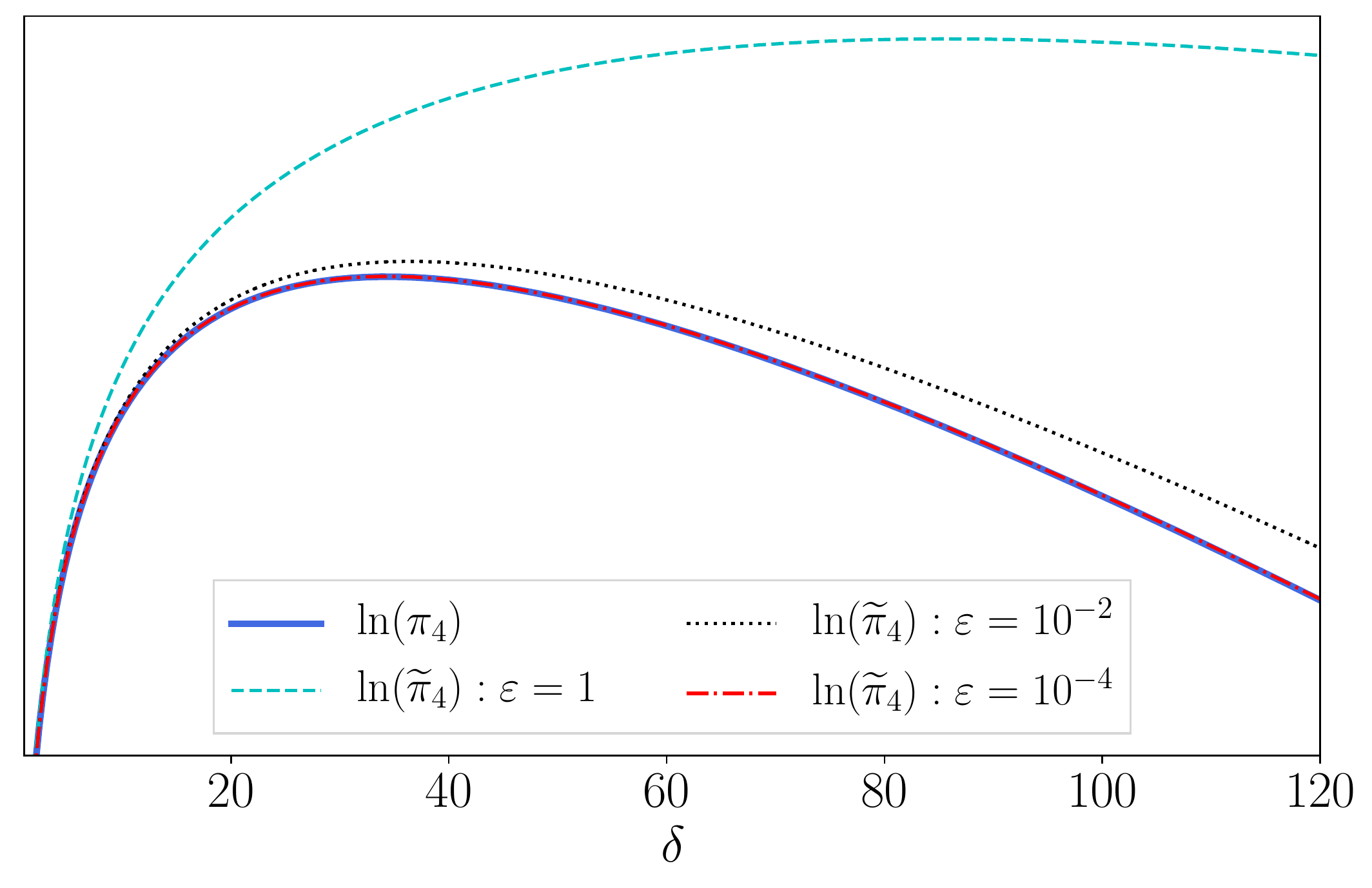}
\caption{Natural logarithm of the exact conditional density $\pi_4$ given in \Cref{eq:cond2} and its approximation $\widetilde{\pi}_4$ given in \Cref{eq:cond_hyper_4a}. Here, $\widetilde{\pi}_4$ depends on the smoothing parameter $\varepsilon$ used to compute the matrix $\mat{L}(\ve{x})$. We show $\ln(\widetilde{\pi}_4)$ for three different values of $\varepsilon$. The vector of attenuation coefficients $\ve{x}$ is assumed known.}
\label{fig:pi4}
\end{figure}

The conditional density $\pi_4$ in \Cref{eq:cond2} can also be represented in terms of a gamma distribution. Using the smooth approximation to the absolute value in the $\ell_1$-norm components of $\pi_4$, we have that $\norm{\mat{D}_1\ve{x}}_1\approx \Vert\mat{W}_1^{\nicefrac{1}{2}}\mat{D}_1\ve{x}\Vert_2^2$ and $\norm{\mat{D}_2\ve{x}}_1\approx \Vert\mat{W}_2^{\nicefrac{1}{2}}\mat{D}_2\ve{x}\Vert_2^2$. This allows us to write the approximation
\begin{equation}
{\pi}_4\left(\delta\given \ve{x}\right) \approx \widetilde{\pi}_4\left(\delta\given \ve{x}\right) \propto \delta^{{d}}\,  \exp\left(-\left[\ve{x}^\tran\mat{L}(\ve{x})\ve{x} +\beta\right]\delta\right), \label{eq:cond_hyper_4a}
\end{equation}
which is proportional to the gamma distribution
\begin{equation}\label{eq:cond_hyper_4}
\widetilde{\pi}_4\left(\delta\given \ve{x}\right)\propto \frac{\beta_{\delta}^{\alpha_{\delta}}}{\Gamma(\alpha_{\delta})}      \delta^{\alpha_{\delta}-1}\exp(-\beta_{\delta}\,\delta), \quad \text{ with } \alpha_{\delta} = d+1 \text{ and }\beta_{\delta}= \ve{x}^\tran\mat{L}(\ve{x})\ve{x} + \beta,
\end{equation}
where $\mat{L}(\ve{x})$ is given in \Cref{eq:L}. \Cref{fig:pi4} shows the natural logarithm of the exact \Cref{eq:cond2} and approximated \Cref{eq:cond_hyper_4a} conditional densities (unnormalized) with different smoothing parameters $\varepsilon$ (for the 50-grain phantom in \Cref{subsec:grains}). They are computed assuming that the true values of the attenuation coefficients $\ve{x}$ are known. In this case, using $\varepsilon<10^{-4}$ yields a close approximation. Hence, instead of sampling the exact conditional \Cref{eq:cond2} via MCMC, we can directly sample from the approximated conditional using \Cref{eq:cond_hyper_4}.

The remaining conditional density $\pi_5$ \Cref{eq:cond3} is one-dimensional and sampling can be achieved using a standard random-walk Metropolis algorithm. Since the parameter $\kappa$ is positive, we apply a logarithmic transformation $\kappa'=\ln(\kappa)$ with the unconstrained parameter $\kappa'$. This parametrization enables the use of random-walk proposals after incorporating the change of variable through a Jacobian in the respective density. Moreover, the Bessel function $I_0(\kappa)$ in \Cref{eq:cond3} is prone to numerical overflow, and thus, we employ the formula $\ln(I_0(\kappa))=\ln({J}_0(\kappa)) + \kappa$, such that the sampler can handle large values of the concentration parameter, here $J_0$ is the exponentially-scaled modified Bessel function of the first kind and order zero. We remark that in contrast to the single-component view angle densities in \Cref{sec:samp_theta}, the scale of the Gaussian proposal can be adapted in this case.

\subsection{The computational procedure}\label{subsec:full}
Based on the sampling schemes discussed in previous subsections, the Gibbs sampler in \Cref{algo:tsGibbs} can be re-written as the hybrid sampler in \Cref{algo:hybrid}. Note that at every iteration, a single sample is drawn from each conditional density. Nevertheless, one can still perform additional $\bar{n}_s$ within-Gibbs iterations when sampling each conditional density to improve the mixing of the final Gibbs chain (cf. \Cref{rem:01}).
\begin{algorithm}[!ht]
\setcounter{AlgoLine}{0}
\DontPrintSemicolon 
\KwIn{conditional densities \Cref{eq:conds}, \yd{finite difference} matrices $\mat{D}_1$ and $\mat{D}_2$, smoothing constant $\varepsilon$, total number of samples $n_s$, number of within-Gibbs samples $\bar{n}_s$, number of CGLS iterations $n_{\mathrm{cgls}}$}

Initial states $\ve{x}^{(0)}, \ve{\theta}^{(0)}, \lambda^{(0)}, \delta^{(0)}, \kappa^{(0)}$\; 

\For{$j = 1,\ldots, n_s$}{

	\texttt{/* Sample attenuation coefficients */}\;

	Solve the least-squares task \Cref{eq:CGLS} with CGLS to obtain a sample of the discretized attenuation coefficient, $\ve{x}^{(j)} \sim \pi_{\mathrm{L}}(\cdot;\ve{x}^{(j-1)})\approx\pi_1(\cdot\given \ve{\theta}^{(j-1)}, \lambda^{(j-1)},\delta^{(j-1)})$ after $n_{\mathrm{cgls}}$ iterations\;

	\texttt{/* Update local Laplace approximation */}\;
	
	Compute weight matrices \Cref{eq:weights}, $\mat{W}_1 = \mat{W}_1(\ve{x}^{(j)})$ and $\mat{W}_2 = \mat{W}_2(\ve{x}^{(j)})$ for the given $\varepsilon$\;
	
	Compute the square root components, $\widetilde{\mat{L}}_{1} =\mat{W}_1^{1/2}\mat{D}_1$ and $\widetilde{\mat{L}}_{2} =\mat{W}_2^{1/2}\mat{D}_2$\;

	Update \yd{the} matrix \Cref{eq:L}, which is equivalent to $\mat{L} =\widetilde{\mat{L}}_{1}^\tran\widetilde{\mat{L}}_{1} +\widetilde{\mat{L}}_{2}^\tran\widetilde{\mat{L}}_{2} $\;

	\texttt{/* Sample view angles */}\;

	Obtain a sample of the view angles, $\ve{\theta}^{(j)} \sim \pi_2(\cdot \given \ve{x}^{(j)},\lambda^{(j-1)}, \kappa^{(j-1)})$ after $\bar{n}_s$ single-component random-walk Metropolis iterations\;

	\texttt{/* Sample hyperparameters */}\;

	Obtain a sample of the precision parameter, ${\lambda}^{(j)} \sim \pi_3(\cdot \given \ve{x}^{(j)}, \ve{\theta}^{(j)})$ using the closed-form \Cref{eq:cond_hyper_3}\;

	Obtain a sample of the inverse scale parameter, ${\delta}^{(j)} \sim \widetilde{\pi}_4(\cdot \given \ve{x}^{(j)})$ using the closed-form \Cref{eq:cond_hyper_4}\;

	Obtain a sample of the concentration parameter, ${\kappa}^{(j)} \sim \pi_5(\cdot \given \ve{\theta}^{(j)})$ after $\bar{n}_s$ random-walk Metropolis iterations\;    
}
\Return{$\{\ve{x}^{(j)}, \ve{\theta}^{(j)}, \lambda^{(j)},\delta^{(j)}, \kappa^{(j)}\}_{j=1}^{n_s}$}\;
\caption{{\sc Hybrid Gibbs sampler to simulate \Cref{eq:target}}}
\label{algo:hybrid}
\end{algorithm}

We measure the efficiency of Algorithm~\ref{algo:hybrid} in terms of the autocorrelation of the hyperparameter chains and the computational cost required for a single iteration. The chain autocorrelation is useful \yd{since it allows} defining the \emph{effective sample size} \cite{owen_2013}
\begin{equation}
n_\mathrm{ESS} := \frac{n_s}{1+2\sum_{j=1}^{n_s}\frac{\rho^{(j)}}{\rho^{(0)}}} \approx \ceil*{\frac{n_s}{\tau_{\mathrm{int}}}},
\end{equation}
where $\rho^{(j)}$ denotes the chain autocorrelation at the $j$th lag and $\tau_{\mathrm{int}}$ is the \emph{integrated autocorrelation time} (IACT). Essentially, the effective sample size is utilized to compare between the variance estimated via correlated MCMC samples and the ideal case of a variance computed from independent draws. Thus, the aim is to obtain an $n_\mathrm{ESS}$ as close as possible to $n_s$. Other relevant metrics utilized in our numerical experiments include analysis of the cumulative mean of the MCMC chains, and the \emph{mean square jump} (MSJ) distance defined as
\begin{equation}
\mathrm{MSJ}:= \frac{1}{N}\sum_{j=1}^{n_s} \norm{\ve{\theta}^{(j)}-\ve{\theta}^{(j-1)}}^2_2,
\end{equation}
here written for the view angle parameter vector $\ve{\theta}$. The MSJ is used as an indicator of how fast the MCMC chains are mixing \cite{martin_et_al_2012}; the larger the magnitude of the MSJ, the better the mixing.

Furthermore, the computational cost of the simulation at each iteration is expressed in terms of the number of CT model calls (forward/backward operations). The cost corresponding to the simulation of each conditional density are: $2{n}_{\mathrm{cgls}}$ calls are needed for $\pi_1$, $\bar{n}_s$ evaluations are required to sample the density $\pi_2$, for the conditional $\pi_3$ one model call is performed, and for $\pi_4$ and $\pi_5$ no evaluations are required. Hence, the total cost of a single iteration of the hybrid sampler is $2n_{\mathrm{cgls}} + \bar{n}_s +1$ model calls.

\section{Numerical experiments}\label{sec:numexp}
Our computational CT model is built on the ASTRA toolbox \cite{vanaarle_et_al_2016}. We employ the line discretization model \Cref{eq:BL_law_discr}; thus, the elements of the system matrix are given by the intersection length of the pixel and the ray of zero thickness.

The projection geometry and data acquisition process comes from a {fan beam} configuration, whose parameters are defined as follows (in arbitrary units): the distance from the X-ray source to the origin is $450$, the distance from the origin to the detector is $150$, the detector length is equal to $300$, the number of detector elements is $p=1.5N$, and the size of the image domain is $N\times N$, where $N$ is chosen in each experiment. The data is generated by adding Gaussian white noise to the forward projection $\mat{A}(\ve{\theta}) \ve{x}_{\mathrm{true}}$ of the true image. The noise standard deviation is ${\sigma}_{\mathrm{obs}}= 0.01 (\Vert\mat{A}(\ve{\theta}) \ve{x}_{\mathrm{true}}\Vert_2/ \sqrt{m})$.

The inputs to hybrid MCMC algorithm are studied/selected as follows: (i) The number of samples is $n_s=2\times 10^4$, in addition to a burn-in period of $n_b=0.2n_s$. The statistics and metrics are computed after thinning the chains by selecting every other component, hence, the final sample size is $10^4$; (ii) The constant controlling the approximation of the $\ell_1$-norm is fixed conservatively to $\varepsilon=10^{-6}$; (iii) The influence of the number of within-Gibbs samples $\bar{n}_s$ and number of CGLS iterations $n_{\mathrm{cgls}}$ is studied in the first example. 

The posterior mean results of the hybrid Gibbs sampler are compared with the reconstructions from the CTVAE algorithm \cite{riis_et_al_2020b}, which is able to account for the view angle uncertainty, but does not provide estimates of the posterior uncertainty and only MAP estimators are computed. The CTVAE solution is only intended for reference with our methodology, rather than for studying which technique performs better. The point estimates $\ve{x}$ of the attenuation coefficients are compared using the relative reconstruction error given by $\eta= \norm{\ve{x}-\ve{x}_{\mathrm{true}}}_2/ \norm{\ve{x}_{\mathrm{true}}}_2$.

\subsection{Grains phantoms}\label{subsec:grains}
Our first example involves the reconstruction of a `{grains}' phantom \cite{hansen_and_jorgensen_2018}. We consider two images with different grain densities; they have $50$ and $100$ grains, respectively.
\begin{figure}[!ht]
\centering
\includegraphics[width=0.99\textwidth]{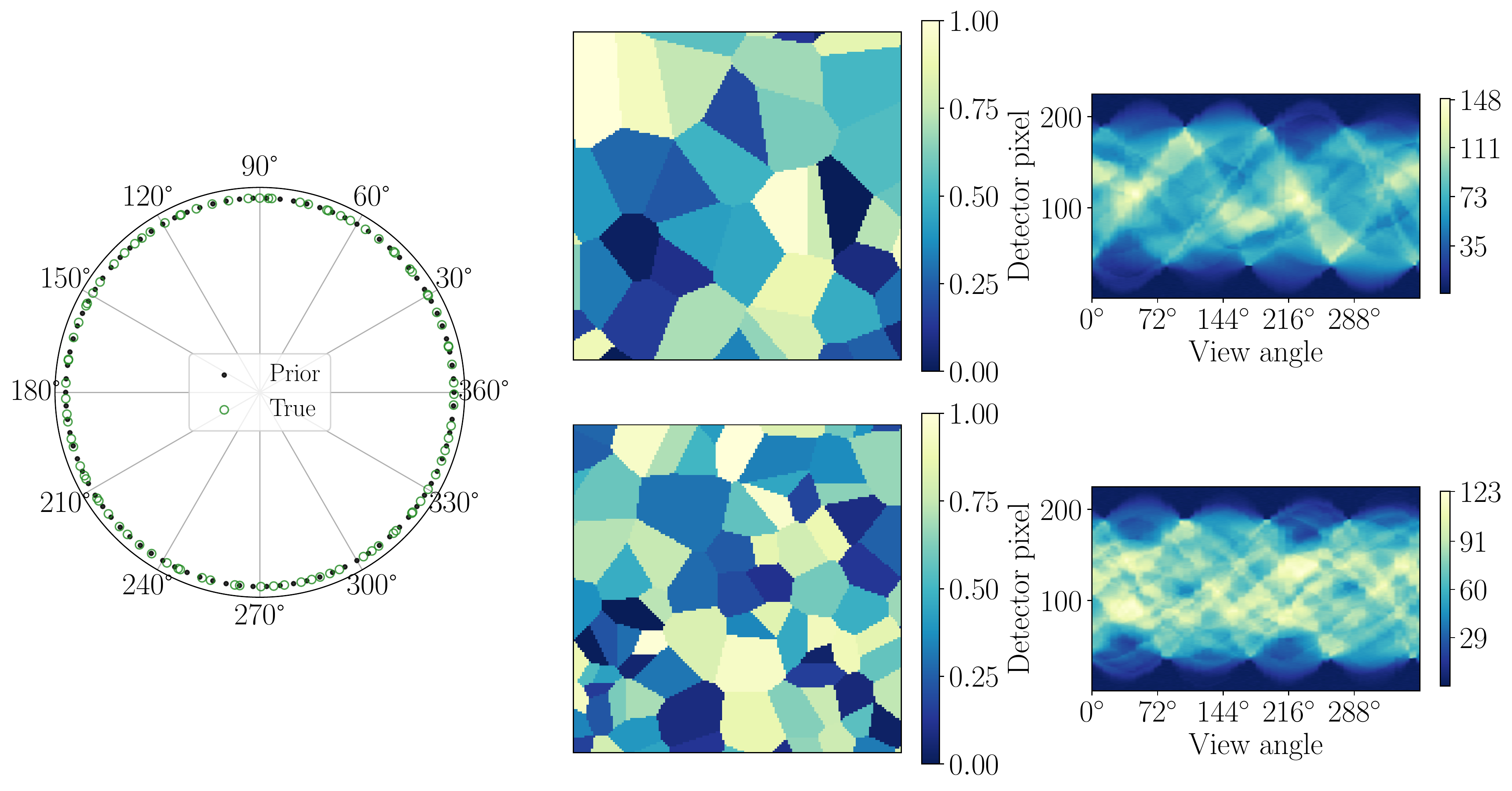}\\
\caption{Grains phantoms with 50 (top) and 100 (bottom) elements. Left: prior mean (nominal) and true view angles. Center: underlying true images. Right: noisy sinogram (data).}
\label{fig:ex_grains_data}
\end{figure}

The discretized domain size is set to $150\times 150$, corresponding to $d=22\,500$ pixels. The set of $q=90$ noisy angle data is assumed to occur during the experiment as a random perturbation of the predefined nominal angles $\ve{a}=[0,4,8,\ldots,356]$. The measurements consist of a vectorized sinogram data $\ve{b}$ with $m=20\,250$ elements. \Cref{fig:ex_grains_data} shows the true images and view angles, together with the sinogram data.

The number within-Gibbs iterations $\bar{n}_s$ utilized in the sampling of the attenuation coefficients is fixed to one (cf. \Cref{rem:02}), and instead we study the number of CGLS iterations $n_{\mathrm{cgls}}$. We perform a parameter study on the value of $\bar{n}_s$ for the view angles and concentration parameters. The idea is to compare the MCMC chain assessment metrics described in \Cref{subsec:full} and posterior statistics (mean and standard deviation). \Cref{tab:metrics} lists these values for the hyperparameters, which indirectly determine the efficiency in sampling the full posterior. We fix $n_{\mathrm{cgls}}= \bar{n}_s= \{5, 10, 20, 40\}$ and employ the phantom with 50 grains. The computational cost per hybrid Gibbs iteration associated to each case is $\{16, 31, 61, 121\}$ forward/backward operations. Note that the posterior statistics remain almost insensitive to the choice of $n_{\mathrm{cgls}}$ and $\bar{n}_s$. This is also reflected in the small changes in the MSJ and IACT, for all hyperparameter chains. Particularly, for the $\lambda$-chain the values of $n_{\mathrm{cgls}}$ and $\bar{n}_s$ have no significant influence, while for the $\delta$ and $\kappa$ chains there are slight improvements when increasing the number of within-Gibbs iterations from 5 to 10. From these results and to keep a tractable computational cost, the number of within-Gibbs iterations is fixed as $n_{\mathrm{cgls}}=\bar{n}_s=10$ in the remainder of the numerical experiments. Moreover, from the IACT and $n_\mathrm{ESS}$ values, we observe that the MCMC samples of the concentration $\kappa$ are almost independent, while roughly half of the inverse scale $\delta$ samples are independent. 
\begin{table}[!ht]
\centering
\caption{Phantom with 50 grains: MCMC chain assessment metrics and posterior statistics for the hyperparameters, with increasing number within-Gibbs iterations  (after burn-in and thinning).}
\label{tab:metrics}
\begin{tabular}{c|c|ccccc}
\hline
Hyperparam. & $n_{\mathrm{cgls}}=\bar{n}_s$ & mean & std & MSJ & IACT & \multicolumn{1}{c}{$n_\mathrm{ESS}$} \\
\hline
\multirow{4}{*}{$\lambda$} & 5 & 2.59  & 0.027 & $1.34\times 10^{-3}$ & 1.29 & $7.70\times10^{3}$ \\
                   & 10 & 2.49 & 0.027 & $1.31\times 10^{-3}$ & 1.33 & $7.52\times10^{3}$ \\
                   & 20 & 2.44 & 0.026 & $1.26\times 10^{-3}$ & 1.23 & $8.16\times10^{3}$  \\
                   & 40 & 2.43 & 0.026 & $1.27\times 10^{-3}$ & 1.27 & $7.88\times10^{3}$  \\
\hline
\multirow{4}{*}{$\delta$} & 5 & 21.90 & 0.169 & $4.30\times 10^{-2}$ & 2.23 & $4.49\times10^{3}$ \\
                   & 10 & 22.51 & 0.174 & $4.76\times 10^{-2}$ & 1.88 & $5.31\times10^{3}$ \\
                   & 20 & 22.80 & 0.178 & $4.84\times 10^{-2}$ & 1.77 & $5.66\times10^{3}$  \\
                   & 40 & 22.80 & 0.180 & $5.04\times 10^{-2}$ & 1.84 & $5.44\times10^{3}$ \\
\hline
\multirow{4}{*}{$\kappa$} & 5 & $1.292\times10^{3}$ & $192.3$ & $7.13\times10^{4}$ & 1.10 & $9.07\times10^{3}$ \\
                   & 10 & $1.287\times10^{3}$ & $188.9$   & $7.12\times10^{4}$ & 1.04 & $9.65\times10^{3}$ \\
                   & 20 & $1.290\times10^{3}$ & $191.7$ & $7.24\times10^{4}$ & 1.09 & $9.17\times10^{3}$  \\
                   & 40 & $1.293\times10^{3}$ & $190.9$ & $7.27\times10^{4}$ & 0.99 & $1.00\times10^{4}$ \\
\hline
\end{tabular}
\end{table}

\begin{figure}[!ht]
\centering
\includegraphics[width=0.99\textwidth]{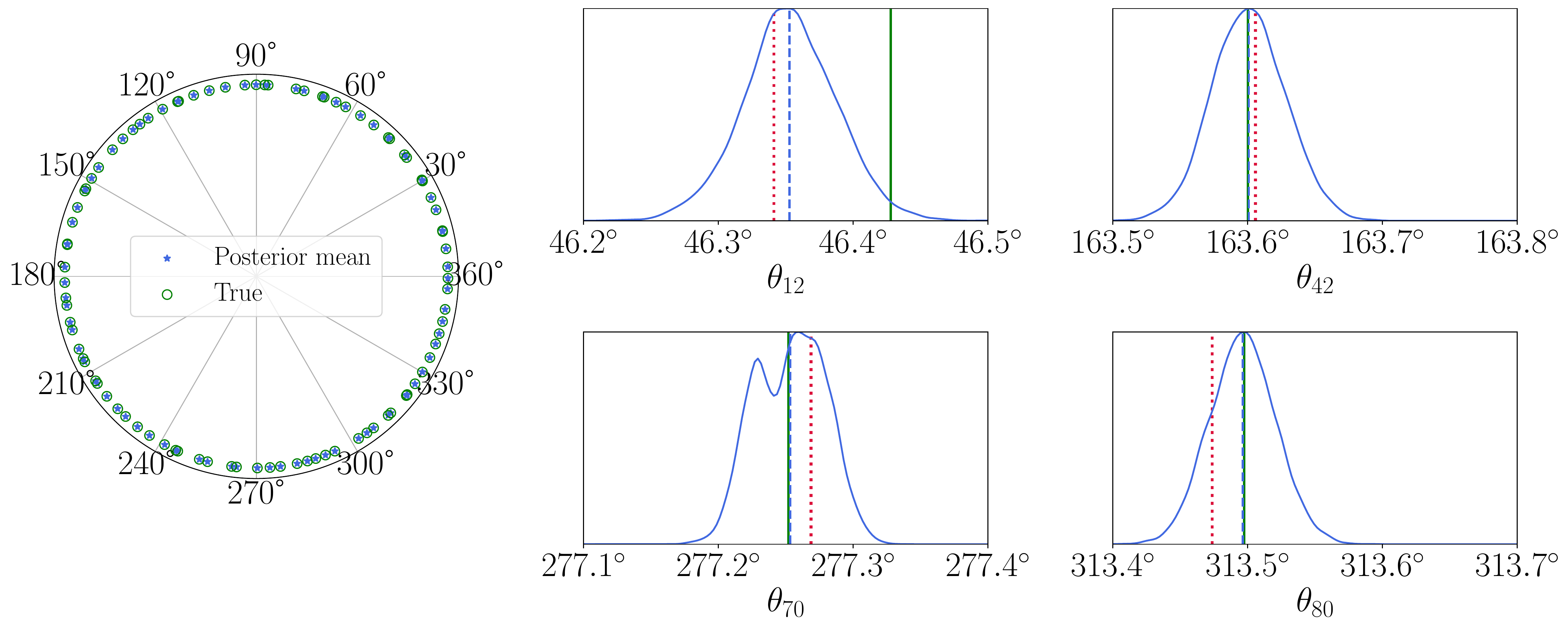}
\caption{Estimated densities of view angles $\theta_i$ with $i=\{12,42,70,80\}$. Left: posterior mean and true view angles. Right: zoom-in on component densities. True angles are shown as solid green lines, posterior mean angles as dashed blue lines, and the angles estimated by the CTVAE algorithm are shown as dotted red lines.}
\label{fig:component_thetas_MCMC}
\end{figure}

The results of the hybrid Gibbs sampler associated to the view angles are shown in \Cref{fig:component_thetas_MCMC}. The proposal spread in the component-wise MCMC algorithm is chosen from the nominal angles as $\sigma=0.05(a_2-a_1)=0.2$ degrees. The posterior mean of the view angles (in blue) is in agreement with the true angles (in green). The MAP solution obtained by the CTVAE algorithm is shown in dotted red lines. The approximated densities are computed from the view angle MCMC chains and we consider $\theta_i$ with $i=\{12,42,70,80\}$. Overall, we observed that some angles are better estimated with CTVAE, while others with hybrid Gibbs. Particularly, for $i=42$ we obtain the best match between the posterior mean and the true view angle; this is opposite to $i=12$ where we obtain the largest deviation from the truth. However, the expected value is still considerably close to the truth given the spread of the densities. Note also that compared to the `true' component densities shown in \Cref{fig:component_thetas}, some of the multimodal characteristics of the density at index 70 are still captured.
\begin{figure}[!ht]
\centering
\hspace*{-0.8cm}\raisebox{-0.2cm}{\small\rotatebox{0}{{CTVAE}}} \hspace*{3.9cm}\raisebox{-0.2cm}{\small\rotatebox{0}{{Hybrid Gibbs}}} \hspace*{3.4cm}\raisebox{-0.2cm}{\small\rotatebox{0}{{Fixed angles}}}\\
\includegraphics[width=0.99\textwidth]{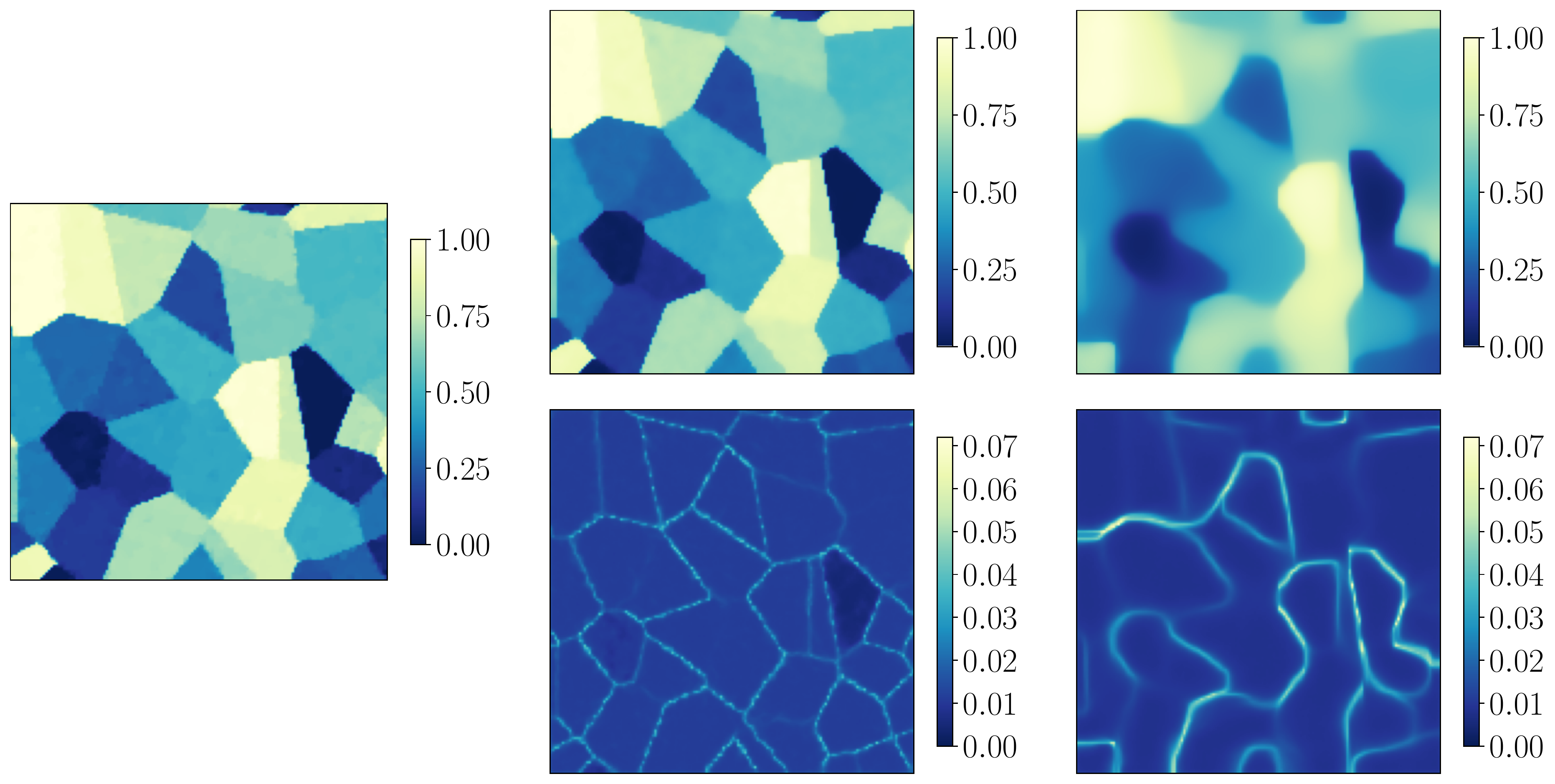}
\caption{Reconstruction of the phantom with 50 grains. We show the MAP estimate of the CTVAE algorithm (first column), and the posterior mean and standard deviation of the proposed hybrid Gibbs sampler (second column), and hybrid Gibbs with fixed nominal angles (third column).}
\label{fig:grains_solution_x1}
\end{figure}
\begin{figure}[!ht]
\centering
\hspace*{-0.8cm}\raisebox{-0.2cm}{\small\rotatebox{0}{{CTVAE}}} \hspace*{3.9cm}\raisebox{-0.2cm}{\small\rotatebox{0}{{Hybrid Gibbs}}} \hspace*{3.4cm}\raisebox{-0.2cm}{\small\rotatebox{0}{{Fixed angles}}}\\
\includegraphics[width=0.99\textwidth]{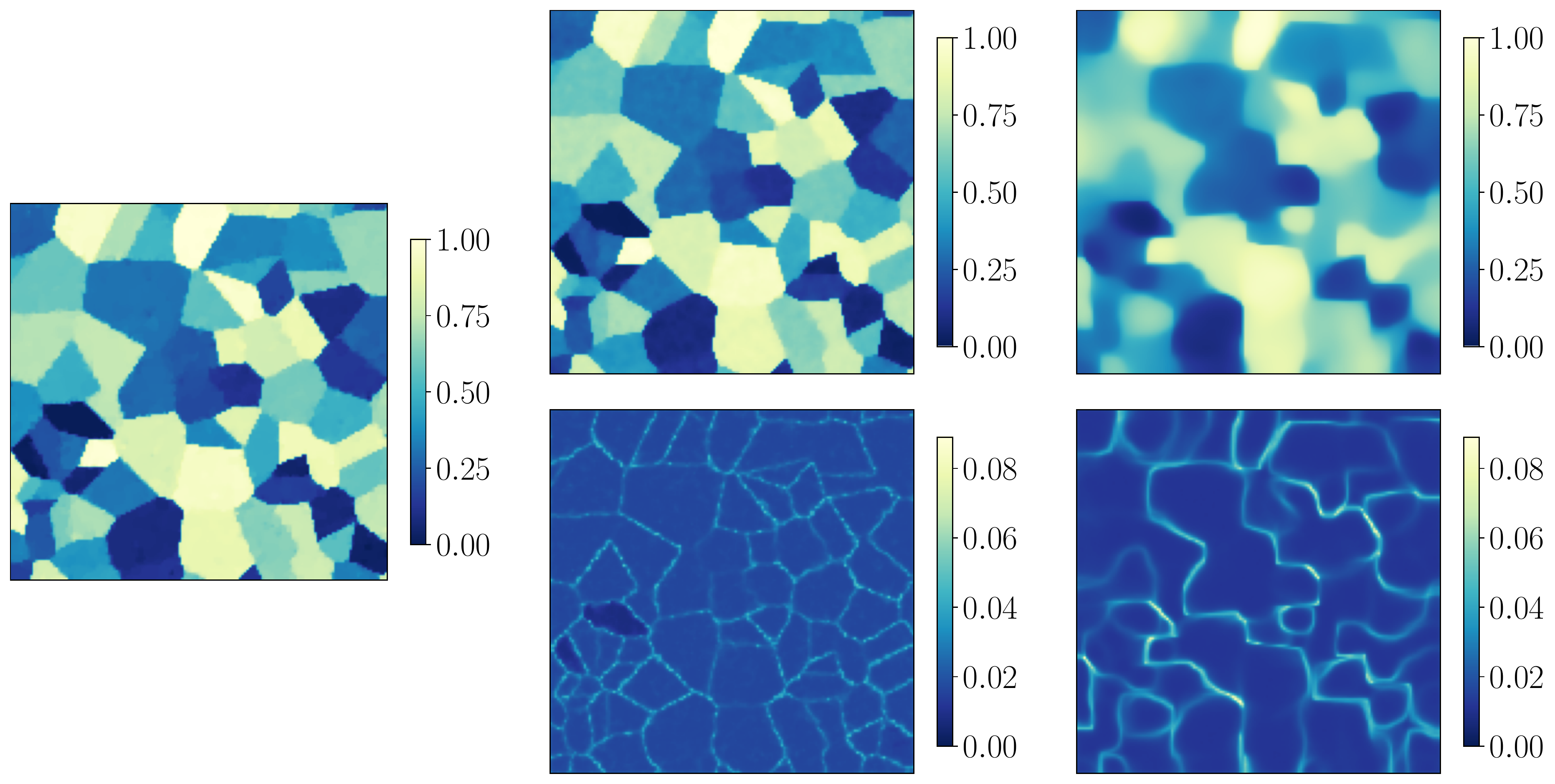}
\caption{Reconstruction of the phantom with 100 grains. We show the MAP estimate of the CTVAE algorithm (first column), and the posterior mean and standard deviation of the proposed hybrid Gibbs sampler (second column), and hybrid Gibbs with fixed nominal angles (third column).}
\label{fig:grains_solution_x2}
\end{figure}

We now focus on the posterior of the discretized attenuation coefficients. For this, we consider two modeling scenarios: (i) hybrid Gibbs with uncertain view angles, and (ii) hybrid Gibbs with fixed nominal angles (i.e., omitting the angle uncertainty). We also compare the posterior mean results with the MAP estimate from the CTVAE method. \Cref{fig:grains_solution_x1} shows the posterior mean and standard deviation for the phantom with 50 grains. The posterior mean of the hybrid Gibbs sampler with uncertain view angles (second column) is in agreement with the underlying true image. Note that it also agrees well the MAP solution computed by CTVAE (first column). The relative reconstruction error of both the posterior mean by hybrid Gibbs and the MAP estimator by CTVAE is $\eta=0.035$. The posterior standard deviation shows that the larger uncertainty is located at the edges of the grains; a result that can be exploited in image segmentation tasks. Conversely, the posterior mean solution obtained by fixing the view angles to the nominal ones (third column) does not yield a satisfactory reconstruction ($\eta=0.15$). In this case, the grains are blurred with no clear distinction of the edges. This is also revealed by the posterior standard deviation where the edges are wider and larger in magnitude.

We perform similar studies with the 100 grains phantom to assess the behavior of the method with an increasing number of grains. In this case, the view angle solution is omitted since we obtain similar results. For the attenuation coefficients, we plot in \Cref{fig:grains_solution_x2} the associated posterior statistics. We consider the same modeling scenarios as in the previous phantom. The relative reconstruction errors for CTVAE, hybrid Gibbs with uncertain angles, and hybrid Gibbs with fixed nominal angles are $\eta=\{0.045, 0.043, 0.19\}$, respectively. The results show that our method also performs well with increasing number of features in the image. Once more, we observe the blurring effect on the posterior mean and larger uncertainty in the edges, when using incorrect angle assumptions.

\subsection{Ppower phantom}
Our second example involves the reconstruction of a `ppower' phantom from \cite{hansen_and_jorgensen_2018}. This is a random image with a zero background and domains in which the pixel values vary smoothly. In this case, we increase the image resolution and reduce the number of projection angles to increase the degree of indeterminacy
(i.e., the problem becomes more under-determined).

\begin{figure}[!ht]
\centering
\includegraphics[width=0.99\textwidth]{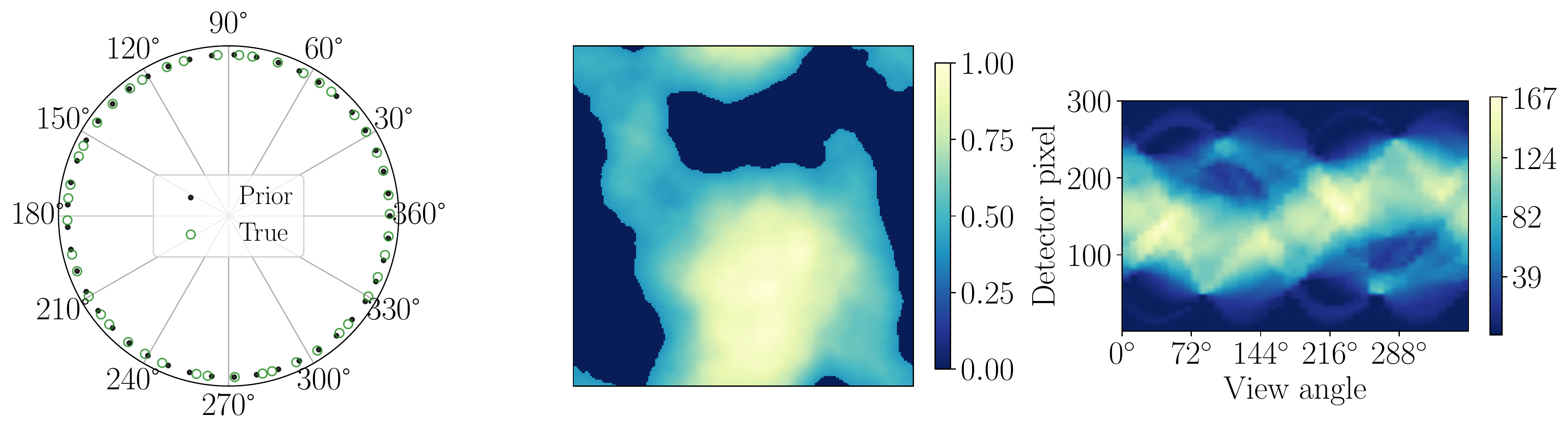}\\
\caption{Ppower phantom. Left: prior and true view angles. Center: underlying true image. Right: noisy sinogram (data).}
\label{fig:ex_power_data}
\end{figure}

The reconstruction domain is discretized with $N = 200$ points in each direction, corresponding to $d=40\,000$ pixels, and we use $q=45$ noisy angles. The nominal angles are given by $\ve{a}=[0,8,\ldots,352]$. The measurements consist of a vectorized sinogram data $\ve{b}$ with $m=13\,500$ elements. \Cref{fig:ex_power_data} shows the true image, the view angles, and the sinogram.

\begin{figure}[!ht]
\centering
\includegraphics[width=0.95\textwidth]{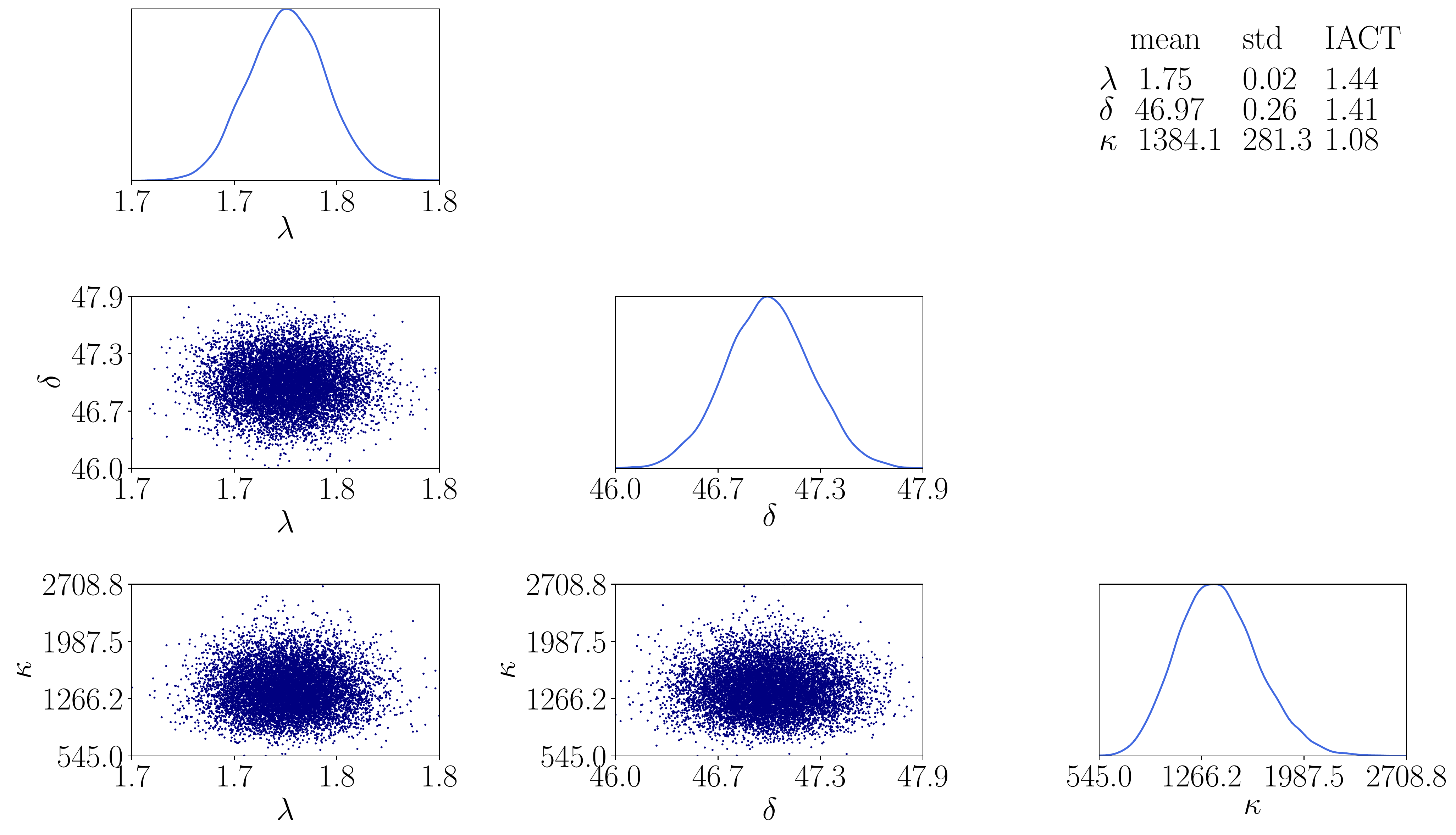}
\caption{Posterior hyperparameters for the ppower phantom: samples and estimated densities. The IACT, together with the posterior mean and standard deviation are also summarized.}
\label{fig:power_hyper}
\end{figure}

\Cref{fig:power_hyper} shows the hyperparameter results after burn-in and thinning. These are given in terms of posterior samples, estimated densities and representative summary statistics, together with the IACT values. Note that the posterior samples for each hyperparameter are relatively independent reflected in IACT values close to one, which also suggest the associated chains are in equilibrium.

\begin{figure}[!ht]
\centering
\includegraphics[width=0.95\textwidth]{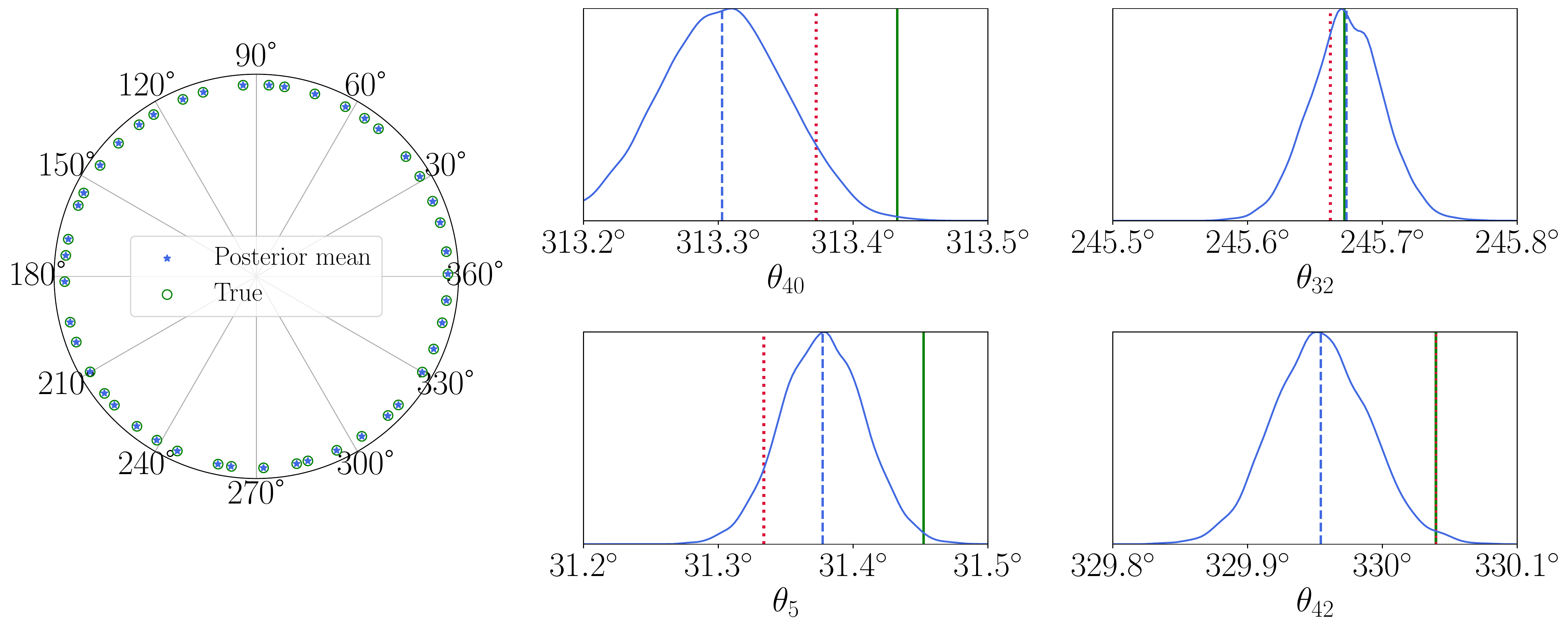}
\caption{Estimated densities of view angles $\theta_i$ with $i=\{40,32,5,42\}$. Left: posterior mean and true view angles. Right: zoom-in on component densities; the top indexes correspond to the largest (left) and smallest (right) deviation between the posterior mean and the true angles; the same results are shown in the bottom indexes for the MAP estimate of the CTVAE. True angles are shown as solid green lines, posterior mean angles as dashed blue lines, and the angles estimated by the CTVAE algorithm as dotted red lines.}
\label{fig:component_thetas_MCMC_power}
\end{figure}

The solution for the view angles is shown in \Cref{fig:component_thetas_MCMC_power}. The proposal scalings in the component-wise MCMC algorithm are chosen from the nominal angles as $\sigma=0.05(a_2-a_1)=0.4$ degrees. The results are illustrated at angle indices $i=\{40,32,5,42\}$. These are selected specifically to show the largest and smallest deviation from the truth computed by point estimates of hybrid Gibbs and CTVAE. Specifically at $i=40$ we obtain the largest difference between the posterior mean estimated by hybrid Gibbs and the true view angle; this compared to $i=32$ where both quantities are the most similar (\Cref{fig:component_thetas_MCMC_power}, top right). For the CTVAE algorithm, the largest and smallest deviation from the truth are given at $i=5$ and 42, respectively (\Cref{fig:component_thetas_MCMC_power}, bottom right). In this case, the solution at three of the shown indexes reveals that the true angle values are located at the tail of the estimated densities. This behavior is only observed for a few angle components, and it can be related to the fixed proposal scaling. Nevertheless, as in the `grains' examples, the view angle component densities are considerably sharp, which makes the posterior mean closer to the true value.

\begin{figure}[!ht]
\centering
\hspace*{-0.8cm}\raisebox{-0.2cm}{\small\rotatebox{0}{{CTVAE}}} \hspace*{3.9cm}\raisebox{-0.2cm}{\small\rotatebox{0}{{Hybrid Gibbs}}} \hspace*{3.4cm}\raisebox{-0.2cm}{\small\rotatebox{0}{{Fixed angles}}}\\
\includegraphics[width=0.99\textwidth]{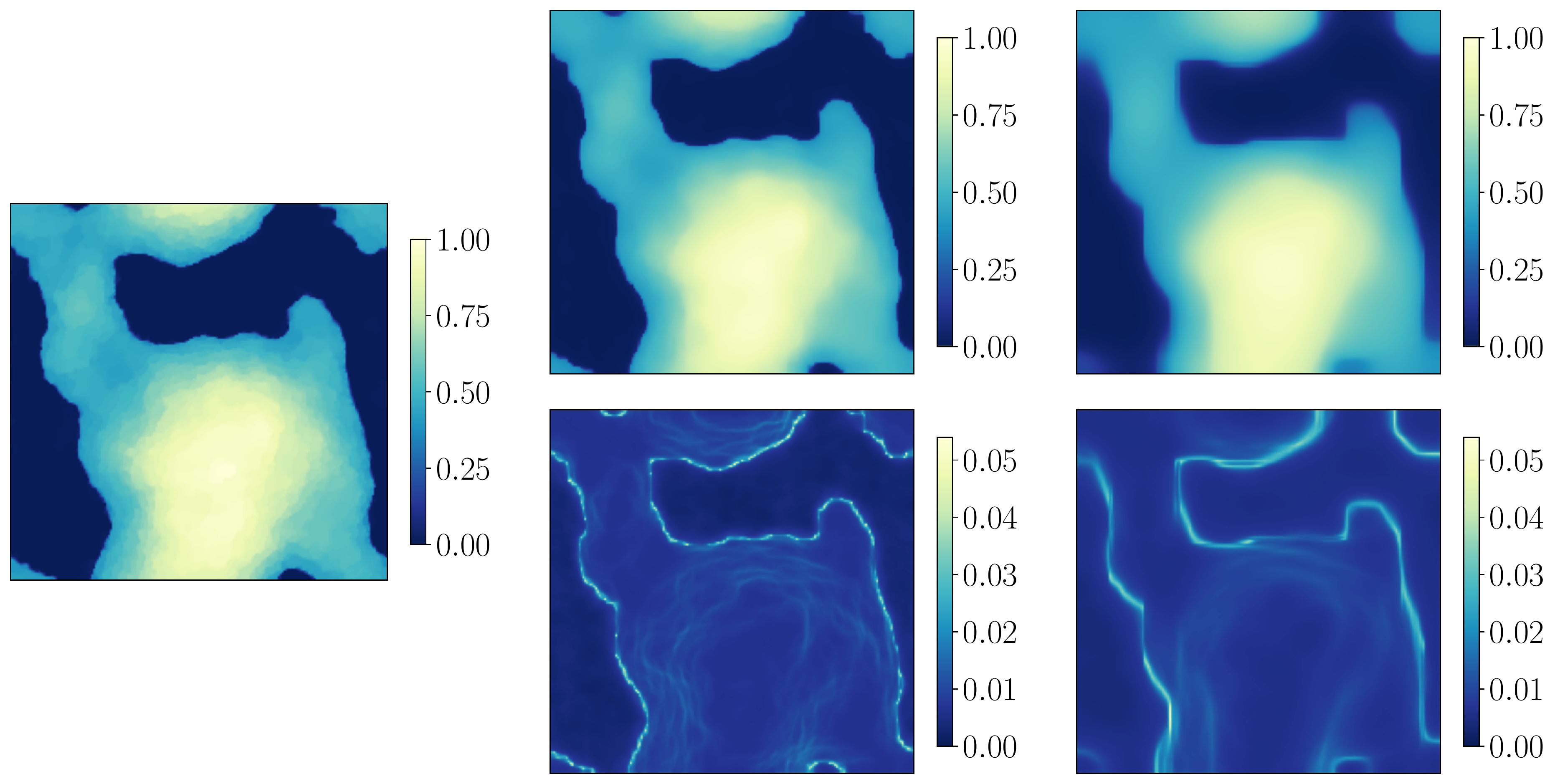}
\caption{Reconstruction of the ppower phantom. We show the MAP estimate of the CTVAE algorithm (first column), the posterior mean and standard deviation of the proposed hybrid Gibbs sampler (second column), and hybrid Gibbs solution with fixed nominal angles (third column).}
\label{fig:power_solution}
\end{figure}

Finally, \Cref{fig:power_solution} shows the reconstructed image. The values of the reconstruction error are $\eta=\{0.035, 0.042, 0.12\}$ for CTVAE (MAP), hybrid Gibbs with uncertain angles (posterior mean), and hybrid Gibbs with fixed nominal angles (posterior mean). Since the CTVAE solution relies on a total variation prior and it computes a MAP estimate, the solution shows more `staircasing' artifacts compared to the posterior mean of our hybrid Gibbs sampler. In this case, one observes that the smooth regions of the image are better represented with the local Laplace approximation used in hybrid Gibbs. Therefore, this representation appears to be a convenient tool in CT reconstruction scenarios where there are not only sharp edges, but also smooth features in the target image. Moreover, the approach that fixes incorrect view angles over-smooths the solution, which shows once more the importance of modeling the view angle uncertainty in X-ray CT applications.

\section{Summary and conclusions}\label{sec:conclusions}
We developed a computational framework for solving high-dimensional hierarchical Bayesian inverse problems. The method is designed for tomographic reconstruction tasks where not only the identification of the image is required, but also the modeling of the uncertainty in the projection angles. The latter aspect is fundamental to correct for potential inaccuracies in the geometry of the experimental setting. Hence, the inverse problem is posed as the joint estimation of the attenuation coefficients representing the image and the view angles characterizing the X-ray projections. We employ a hierarchical regularization strategy where different hyperparameters are also incorporated into the inverse problem. The resulting joint posterior distribution is characterized using the Gibbs sampler. Each conditional density within the Gibbs structure defines individual Bayesian inverse problems, which are solved with different sampling schemes (thereby defining a hybrid Gibbs sampler).

In order to sample the conditional associated with the high-dimensional attenuation coefficients, we propose a method based on a local Laplace approximation. The main idea is to employ this representation as a proposal distribution for MCMC techniques that can be used to adjust the samples to follow the underlying target conditional. In our formulation, the Gaussian proposal is sampled efficiently with an iterative least squares solver. We discussed that sampling from the unadjusted Laplace approximation within the Gibbs iterations is equivalent to an independence sampler MCMC setting with one within-Gibbs iteration. Our numerical experiments show that this approach is sufficient to obtain satisfactory image reconstructions and posterior uncertainty estimates, while keeping a tractable computational cost. Furthermore, the conditional associated with the view angles is characterized with a component-wise Metropolis algorithm. We observed that the single-component view angle conditionals are very sharp and some of them are multimodal. Our component-wise formulation benefits from the fact that the forward projections in the computed tomography model can be readily computed individually for each individual angle. Despite its simplicity, the results illustrate that the single-component sampler is able to simulate the underlying view angles. They also reveal that the solution has potential to be improved by considering MCMC techniques that perform well for sharp distribution and the application of proposal adaptation steps.

Our approach is suitable in cases when preserving the edges of the image is also an important aspect of the reconstruction. In fact, our method is tailored to Bayesian inverse problems that utilize Laplace difference priors on the attenuation coefficients. An advantage of this formulation is that the edges are revealed in the posterior uncertainty\,---\,the largest variability is concentrated at the edges. As a result, statistics of the posterior attenuation coefficients, such as the standard deviation, can be exploited in image segmentation tasks.

Finally, some ideas to extend/improve our methodology are as follows. (i) When other edge-preserving priors are chosen for the attenuation coefficients, the MCMC adjustment step to our Laplace approximation becomes essential to obtain good approximations to the target conditional. Hence, the development of efficient computational techniques that enable the practical application of MCMC adjustment is necessary, e.g., low-rank approximations suitable for heavy-tailed prior distributions. (ii) The consideration of a learning rate parameter in the quasi-Newton step can potentially improve the sampling of the attenuation coefficients in cases where the target conditional differs significantly from a Gaussian distribution. (iii) The single-component view angle conditionals are sharp and some of them are multimodal. The application of MCMC techniques tailored for this type of problems and the implementation of adaptation steps can help to obtain more accurate view angle estimates.

\section*{Acknowledgements}\label{sec:acknowledgements}
This work was supported by a Villum Investigator grant (no.\ 25893) from The Villum Foundation.

\addcontentsline{toc}{section}{References}
\bibliographystyle{model1-num-names}
\bibliography{bibfile.bib}

\end{document}